\documentclass[
    aps,prd,
    twocolumn,
    nofootinbib,
    superscriptaddress,
]{revtex4}
\usepackage[utf8]{inputenc}
\usepackage{lmodern}
\usepackage{amsmath}
\usepackage{amssymb}
\usepackage{makecell}
\usepackage{amsfonts,bm}
\usepackage{graphicx}
\usepackage{float}
\usepackage{paralist}
\usepackage[colorlinks]{hyperref}

\begin{document}

\title{Sensitivities to feebly interacting particles: public and unified calculations}

\author{Maksym~Ovchynnikov}
\email{maksym.ovchynnikov@kit.edu}
\affiliation{Institut für Astroteilchen Physik, Karlsruher Institut für Technologie (KIT), Hermann-von-Helmholtz-Platz 1, 76344 Eggenstein-Leopoldshafen, Germany}
\affiliation{Instituut-Lorentz, Leiden University, Niels Bohrweg 2, 2333 CA Leiden, The Netherlands}

\author{Jean-Loup~Tastet}
\email{jean-loup.tastet@uam.es}
\affiliation{Departamento de Física Teórica and Instituto de Física Teórica UAM/CSIC,
Universidad Autónoma de Madrid, Cantoblanco, 28049, Madrid, Spain}

\author{Oleksii~Mikulenko}
\email{mikulenko@lorentz.leidenuniv.nl}
\affiliation{Instituut-Lorentz, Leiden University, Niels Bohrweg 2, 2333 CA Leiden, The Netherlands}

\author{Kyrylo~Bondarenko}
\email{kyrylo.bondarenko@sissa.it}
\affiliation{IFPU, Institute for Fundamental Physics of the Universe, via Beirut 2, I-34014 Trieste, Italy}
\affiliation{SISSA, via Bonomea 265, I-34132 Trieste, Italy}
\affiliation{INFN, Sezione di Trieste, SISSA, Via Bonomea 265, 34136, Trieste, Italy}

\date{\today}

\begin{abstract}
The idea that new physics could take the form of feebly interacting particles (FIPs) --- particles with a mass below the electroweak scale, but which may have evaded detection due to their tiny couplings or very long lifetime --- has gained a lot of traction in the last decade, and numerous experiments have been proposed to search for such particles. It is important, and now very timely, to consistently compare the potential of these experiments for exploring the parameter space of various well-motivated FIPs. The present paper addresses this pressing issue by presenting an open-source tool to estimate the sensitivity of many experiments --- located at Fermilab or the CERN's SPS, LHC, and FCC-hh --- to various models of FIPs in a unified way: the \texttt{Mathematica}-based code \texttt{SensCalc}.
\end{abstract}

\maketitle

\section{Introduction}
\label{sec:introduction}

The well-known shortcomings of the Standard Model (SM) suggest to us the existence of new physics ``Beyond the Standard Model'' (BSM), which is generally expected to involve new particles. There is currently no clear theoretical guidance, nor experimental hints, about the mass of the hypothetical new particles, which could range from sub-$\text{eV}$ all the way up to the Planck scale. Particles with a mass below the electroweak scale are of particular interest since they may be numerously produced at accelerators. The past experiments have already pushed the limits on the couplings of such particles to tiny values; hence they are called \textit{feebly interacting particles}, or \textit{FIPs} for short. FIPs may be searched for at the main detectors of colliders (ATLAS, CMS, ALICE, and LHCb at the LHC, or their equivalents at future colliders such as the FCC-hh) which are located very close to the collision point, or at so-called \textit{lifetime-frontier} experiments, which re-use existing facilities or infrastructure and place a decay volume near the interaction point or target. Lifetime-frontier experiments may be broadly split into two classes~\cite{Beacham:2019nyx}: collider-based, which make use of the interaction points of ATLAS, CMS, and LHCb, and extracted-beam experiments, which use an extracted beam line hitting a target. 

During the last few years, many lifetime-frontier experiments have been proposed. Among extracted-beam experiments, we can list SHiP~\cite{Alekhin:2015byh,Aberle:2839677}, SHADOWS~\cite{Alviggi:2839484}, and HIKE$_{\text{dump}}$~\cite{CortinaGil:2839661} at the SPS, and DUNE~\cite{DUNE:2020ypp,DUNE:2020fgq} and DarkQuest~\cite{Batell:2020vqn} at Fermilab. The proposed LHC-based experiments include MATHUSLA~\cite{MATHUSLA:2019qpy} and FACET~\cite{Cerci:2021nlb}, associated with CMS; FASER~\cite{FASER:2018bac}, SND@LHC~\cite{SHiP:2020sos} (together with their upgrades, AdvSND and FASER2) and ANUBIS~\cite{Bauer:2019vqk}, close to the ATLAS interaction point; CODEX-b~\cite{Aielli:2019ivi} near LHCb; and AL3X~\cite{Dercks:2018wum} at ALICE. Furthermore, lifetime-frontier experiments will likely remain a part of the physics program of future colliders, such as the {FCC-hh} \cite{Boyarsky:2022epg}.

To evaluate the potential of those experiments to search for generic FIPs, the Physics Beyond Colliders (PBC) initiative has proposed~\cite{Beacham:2019nyx} a few benchmark models. They include dark photons, millicharged particles, dark scalars, heavy neutral leptons, and axion-like particles coupled to various SM particles.

While some of the experiments from the above list are already running, many are still at the status of proposals. Their design is not finalized yet and is still undergoing optimization. Their sensitivity can be optimized by focusing on two key aspects: increasing the rate of events with FIPs while reducing the SM backgrounds. Studying the background requires knowing the detailed specifications of the experimental setup, background-reducing systems, and surrounding infrastructure. As a result, full simulations are required, which accurately trace each event, starting from the initial proton collision and ending with the interactions of the background particles with the detector material. Most of the experimental proposals claim to achieve zero background level. In contrast, the evaluation of the FIP event rate is comparatively less affected by these complexities. This is the case, in particular, when the FIPs are produced at the collision point. They would then propagate through the infrastructure without being affected by the material (due to their tiny interaction strength) and decay or scatter inside the decay volume with some tiny probability. If the reaction products reach the detector (and satisfy some simple kinematic cuts), they typically can be detected with $\approx 1$ efficiency. Therefore, the sensitivity\footnote{When talking about ``sensitivity'', it is important to point out the distinction between ``exclusion'' sensitivity (rejecting the New Physics hypothesis in the absence of signal) and ``discovery'' sensitivity (rejecting the Standard Model in favor of New Physics if a signal is observed). While the former is not very sensitive to the exact background expectation as long as it is $\lesssim 1$, the latter strongly depends on it. Throughout this paper, we mean ``exclusion sensitivity'' whenever we use the word ``sensitivity'' unqualified. However, we advise the reader to keep this distinction in mind when comparing the physics potential of various experiments: indeed, two experiments with the same exclusion sensitivity could, in principle, have significantly different discovery sensitivities.} of a given experiment to FIPs is determined mainly by 1) the distribution of FIPs at the facility housing the experiment and 2) the geometry of the experiment itself.

Despite the relative simplicity of estimating the sensitivity to FIPs, a few caveats can make it challenging to compare different experiments. First, there is often no unique description of the production and decay of a given FIP in the literature. This is related to either theoretical uncertainties in the description of the FIP phenomenology or different conventions in the definition of the model. As a result, different experimental collaborations can end up using different descriptions of the FIPs; sometimes, even the definition of the FIP coupling is different (see Appendix~\ref{app:couplings-definitions}). Secondly, due to the rapid pace of change as the experiment's design is being optimized, there may exist a mismatch between, on the one hand, the experimental setup and/or the assumptions used and, on the other hand, the reported sensitivity, even within a same document (see Fig.~\ref{fig:validation-off-axis} and the corresponding discussion). Indeed, to update the sensitivity while the setup is undergoing optimization, collaborations would need to re-launch full-scale simulations, which require a lot of time, computational resources, and person-power. Third, the collaborations' tools for performing sensitivity calculations are typically ``black boxes'' for outsiders, since they are not publicly accessible. As a result, they do not provide a qualitative understanding of the sensitivity and thus prevent simple cross-checking against errors or numerical artifacts. This problem becomes especially important when comparing the sensitivities of various experiments to understand which one is better suited to probe a given region of the FIP parameter space.

To address these issues, a public tool that can calculate the sensitivity of various experiments to FIPs in a unified and transparent way is required. Several publicly available packages can already perform such sensitivity calculations~\cite{Kling:2021fwx,Jerhot:2022chi}. However, they are limited to a specific type of facilities: either beam dump experiments or colliders. This paper presents the \texttt{Mathematica}~\cite{Mathematica} code\footnote{Although the \texttt{SensCalc} package is technically open-source, it requires a copy of the Mathematica software to run.} \texttt{SensCalc}~\cite{SensCalc-Zenodo} that can evaluate the sensitivity of the various experiments proposed at Fermilab, SPS, LHC, and FCC-hh to various models of FIPs.\footnote{Available at \url{https://doi.org/10.5281/zenodo.7957784}\, and also at \url{https://github.com/maksymovchynnikov/SensCalc}\,.} The code is based on a semi-analytic approach developed in Ref.~\cite{Bondarenko:2019yob}, and further improved and cross-checked in Refs.~\cite{Boiarska:2021yho,Boyarsky:2022epg,Ovchynnikov:2022its} (see also~\cite{Coloma:2023adi,Batell:2023mdn}). The number of events is approximated by the integral of several quantities: the FIP angle-energy distribution, decay probability, geometric acceptance, and the acceptance of its decay products. Most of these quantities can be accurately computed analytically, which is especially attractive as it improves the transparency of the computations.

The present paper is organized as follows. In Sec.~\ref{sec:approach}, we discuss the semi-analytic method we use to calculate the sensitivity, along with its validation and limitations. In Sec.~\ref{sec:senscalc}, we provide a brief description of \texttt{SensCalc}, specifying the list of the currently implemented experiments and models of FIPs. We also compare it with other publicly available packages for computing the sensitivity, as well as with \texttt{SensMC}~\cite{SensMC-GitHub}, a simplified Monte-Carlo simulation that we have specifically developed to validate it. In Sec.~\ref{sec:case-studies}, we demonstrate two use cases for \texttt{SensCalc}. Finally, we conclude in Sec.~\ref{sec:conclusions}.
In the appendices, we expand on a number of topics that we kept out of the main text in the interest of brevity. In App.~\ref{app:couplings-definitions}, we discuss various discrepancies present in the literature stemming from different descriptions of the interactions or conventions for the couplings. In App.~\ref{app:references}, we detail the various inputs we use to compute the signal yields. Finally, in App.~\ref{app:toy-mc}, we briefly describe the operation of \texttt{SensMC}.

\section{Semi-analytic approach to calculate sensitivities}
\label{sec:approach}

\subsection{Method}
\label{sec:method}

This work concentrates on FIPs produced directly at the collision point or in its immediate vicinity. In this case, the production is unaffected by the surrounding infrastructure. We calculate the number of events involving a decaying FIP using the following expression: 
\begin{multline}
    N_{\text{ev}} = \sum_{i}N^{(i)}_{\text{prod}}\int dE d\theta dz \ f^{(i)}(\theta,E)\cdot \epsilon_{\text{az}}(\theta,z) \cdot \\
    \cdot \frac{dP_{\text{dec}}}{dz} \cdot\epsilon_{\text{dec}}(m,\theta,E,z) \cdot \epsilon_{\text{rec}}
    \label{eq:Nevents}
\end{multline}

\begin{figure}[!h]
    \centering
    \includegraphics[width=\linewidth]{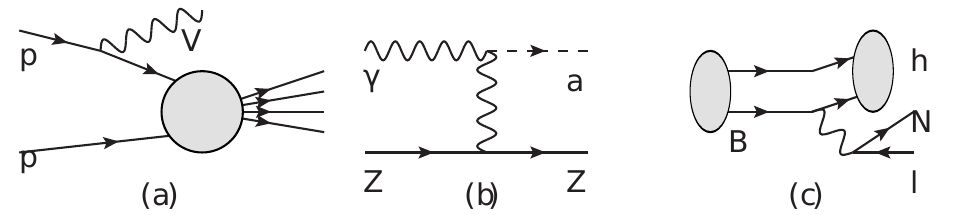}
    \caption{Examples of production processes for various FIPs: (a) proton bremsstrahlung (for the dark photon $V$), (b) coherent scattering off nuclei (for the ALP $a$ coupling to photons), (c) decays of $B$ mesons into a FIP and another meson $h$ (for HNLs $N$).}
    \label{fig:fip-production-channels}
\end{figure}

The quantities entering Eq.~\eqref{eq:Nevents} are the following:
\begin{itemize}
\item[--] $N_{\text{prod}}^{(i)}$ is the total number of FIPs produced by the process $i$, e.g., decays of mesons, direct production by proton-target collisions, etc.\ (see fig.~\ref{fig:fip-production-channels}).
\item[--] $z$, $\theta$, and $E$ are, respectively, the position along the beam axis, the polar angle, and the energy of the FIP.
\item[--] $f^{(i)}(\theta,E)$ is the differential distribution of FIPs in polar angle and energy for FIPs produced through the process~$i$.
\item[--]$\epsilon_{\text{az}}(\theta,z)$ is the azimuthal acceptance: 
    \begin{equation}
    \epsilon_{\text{az}} = \frac{\Delta \phi_{\text{decay volume}}(\theta,z)}{2\pi}
    \end{equation}
where $\Delta\phi$ is the fraction of azimuthal coverage for which FIPs decaying at $(z,\theta)$ are inside the decay volume.
\item[--] $\frac{dP_{\text{dec}}}{dz}$ is the differential decay probability: 
\begin{equation}
\frac{dP_{\text{dec}}}{dz} = \frac{\exp[-r(z,\theta)/l_{\text{dec}}]}{l_{\text{dec}}} \frac{dr(z,\theta)}{dz},
\end{equation}
with $r = z/\cos(\theta)$ being the modulus of the displacement of the FIP decay position from its production point, and $l_{\text{dec}}=c\tau \sqrt{\gamma^2-1}$ is the FIP decay length in the lab frame.
\item[--] $\epsilon_{\text{dec}}(m,\theta,E,z)$ is the decay products acceptance, i.e.\ among those FIPs that are within the azimuthal acceptance, the fraction of FIPs that have at least two decay products that point to the detector and that may be reconstructed. Schematically,
\begin{equation}
    \epsilon_{\text{dec}} = \text{Br}_{\text{vis}}(m)\cdot \epsilon_{\text{dec}}^{\text{(geom)}}\cdot \epsilon_{\text{dec}}^{\text{(other cuts)}}
    \label{eq:decay-acceptance-schematic}
\end{equation}
Here, $\text{Br}_{\text{vis}}$ denotes the branching ratio of the FIP decays into final states that are detectable; depending on the presence of a calorimeter (EM and/or hadronic), $\text{Br}_{\text{vis}}$ may encompass only those states featuring at least two charged particles, or it may also include some neutral states such as photons and $K^{0}_{L}$. $\epsilon_{\text{dec}}^{\text{(geom)}}$ denotes the fraction of visible decay products that point to the end of the detector, and $\epsilon_{\text{dec}}^{\text{(other cuts)}}$ is the fraction of these decay products that additionally satisfy the remaining cuts (e.g., the energy cut, etc.).
\item[--]$\epsilon_{\text{rec}}$ is the reconstruction efficiency, i.e., the fraction of the events that pass the azimuthal and decay acceptances criteria that the detector can successfully reconstruct. It results from the non-ideal performance of the detector, which introduces a finite detection efficiency and kinematics measurement resolution.
\end{itemize}

\begin{figure*}
    \centering
    \includegraphics{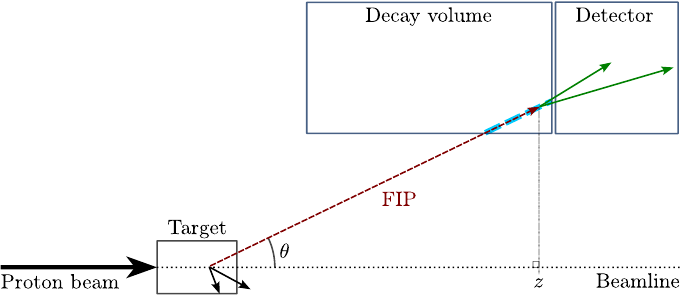}
    \caption{Schematic diagram of a beam dump experiment with a detector located downstream of the decay volume (cf.\ text for details).}
    \label{fig:acceptance-illustration}
\end{figure*}

Fig.~\ref{fig:acceptance-illustration} illustrates the impact of the different contributions on the number of events~\eqref{eq:Nevents}. Consider a FIP decaying at coordinates $(\theta,z)$, where $\theta$ is the polar angle relative to the beamline, $z$ is the longitudinal displacement from the target, and the azimuthal angle $\phi$ has been omitted from the diagram. The differential probability for a FIP with energy $E$ to decay there is $f(\theta,E)dP_{\text{dec}}/dz$. The azimuthal coordinate~$\phi$ of the decaying FIP (whose trajectory is shown by the red arrow) must be within the decay volume, which restricts the available decay positions to the blue dashed line. These limitations are included in the azimuthal acceptance $\epsilon_{\text{az}}$. Next, at least two of the FIP decay products (the green arrows) have to point to the detector; this is accounted for in $\epsilon_{\text{dec}}$. Depending on the setup and the FIP, this requirement may significantly limit the decay volume's ``useful'' angular coverage. In particular, for 2-body decays into stable particles, the decay products can only point to the detector if the decayed FIP also points to the detector. Only the narrow angular domain that the detector covers contributes to the number of events.

Most quantities entering Eq.~\eqref{eq:Nevents} can be accurately estimated analytically and cross-checked separately, which makes the approach~\eqref{eq:Nevents} very transparent. Namely, the azimuthal acceptance is completely determined by the geometry of the decay volume, which is typically very simple. Once $\epsilon_{\text{az}}$ is computed, a simple way to cross-check it is to verify that the integral
\begin{equation}
\begin{aligned}
    \mathcal{V} &= 2\pi \int d\theta dr r^{2}(z,\theta)\sin(\theta) \epsilon_{\text{az}} \\
    &= 2\pi \int d\theta dz \frac{z^{2}}{\cos^{3}(\theta)}\sin(\theta)\epsilon_{\text{az}}
\end{aligned}
\end{equation}
matches the total volume of the decay volume.

Depending on the production channel, evaluating the FIP distribution function $f^{(i)}(\theta,E)$ may require some external input. For instance, for FIPs that are produced directly in inelastic proton collisions, one needs to simulate $f^{(i)}(\theta,E)$ using, e.g.,\ \texttt{PYTHIA~8} to account for showering and hadronization. For FIPs that are produced in the interactions of secondary particles, either in their decays or scattering with the material (see Fig.~\ref{fig:fip-production-channels} for examples), the distribution of secondaries $f_{\text{secondary}}(\theta,E)$ is needed; nevertheless, once $f_{\text{secondary}}(\theta,E)$ has been computed, the distribution of FIPs can then be derived analytically without the need for external tools.

The decay acceptance $\epsilon_{\text{dec}}$ may, in principle, be estimated qualitatively by comparing the opening angle $\Delta \theta_{\text{dec}}$ between the decay products with the angle $\Delta \theta_{\text{det}}$ covered by the detector as seen from the production point. In the simplest case of a two-body decay into massless particles, the opening angle is $\Delta \theta_{\text{dec}} \simeq 2\arcsin(\gamma^{-1})$, where $\gamma$ is the boost factor of the FIP. If the detector is too small to cover such an angle, $\Delta \theta_{\text{dec}} \gtrsim \Delta \theta_{\text{det}}$,  it would have a low sensitivity $\epsilon_{\text{dec}}\approx 0$, otherwise $\epsilon_{\text{dec}}\approx 1$. Because the detector angle is smallest at the beginning of the decay volume while the opening angle decreases as $E_{\text{FIP}}^{-1}$, $\epsilon_{\text{dec}}$ effectively imposes a cut from below on the FIP energy and the displacement of its decay position from the beginning of the decay volume. If the detector itself constitutes the decay volume (as in the case of, e.g., neutrino detectors), then the decay products are being tracked directly from the decay vertex and $\epsilon_{\text{dec}}^{\text{(geom)}}\equiv 1$.

To estimate $\epsilon_{\text{dec}}$ more accurately, by accounting for such factors as the experiment geometry, the presence of a dipole magnet, different FIP decay topologies (such as multi-body decays or decays into unstable particles), and various other selections imposed on the decay products, one can perform a \emph{separate} simulation (see details in Sec.~\ref{sec:senscalc}).

Finally, the computation of $\epsilon_{\text{rec}}$ would require running the full simulation, including the detector response. As such, it goes beyond the scope of the present semi-analytic approach. However, we believe that it is possible to perform an adequate pre-selection with the help of $\epsilon_{\text{dec}}^{\text{(other cuts)}}$ (for instance, by requiring the energy or $p_{\mathrm{T}}$ of the final state particles to exceed a threshold above which they are detected with high efficiency; see, e.g.,~\cite{Aberle:2839677}), such that, conditioned on this pre-selection, $\epsilon_{\text{rec}} \sim \mathcal{O}(1)$. In addition, pre-computed reconstruction efficiencies (for instance, the reconstruction efficiency as the function of the track's energy for the given particle type) may be available.

Last but not least, this semi-analytic method allows for a simple analysis of the number of events in the limit of very long-lived FIPs with lifetimes $c\tau\langle \sqrt{\gamma^{2}-1}\rangle\gg l_{\text{experiment}}$, where $l_{\text{experiment}}$ is the length scale of the experiment. In this case, the only dependence of the number of events~\eqref{eq:Nevents} on $c\tau$ is multiplicative:
\begin{equation}
    N_{\text{ev}} \approx \sum_{i}N_{\text{prod}}^{(i)}\cdot \epsilon^{(i)},
    \label{eq:number-of-events-large-lifetimes}
\end{equation}
where $\epsilon^{(i)}$ is the total acceptance for the given production channel:
\begin{equation}
    \epsilon^{(i)} = \int d\theta dE dz\ f^{(i)}\cdot \epsilon_{\text{az}}\cdot \frac{\epsilon_{\text{dec}}}{\cos(\theta)c\tau\sqrt{\gamma^{2}-1}},
    \label{eq:total-acceptance}
\end{equation}
and the function parameters have been omitted for brevity.
This quantity may be decomposed as
\begin{equation}
    \epsilon = \langle \epsilon_{\text{FIP}}\rangle\cdot \langle (\gamma^{2}-1)^{-1/2}\rangle\cdot \frac{\Delta z}{c\tau}\cdot \langle \epsilon_{\text{decay}}\rangle,
\end{equation}
where $\langle \epsilon_{\text{FIP}}\rangle$ is the mean probability for the FIP to intersect the decay volume, $\langle (\gamma^{2}-1)^{-1/2}\rangle$ is the mean inverse $p/m$ among the FIPs meeting the azimuthal criterion, $\Delta z$ is the total longitudinal size of the decay volume, and $\langle \epsilon_{\text{decay}}\rangle$ is the mean decay products acceptance. This representation is particularly useful when discussing the impact of the geometry on the event rate and comparing the potential of various experimental setups~\cite{Bondarenko:2019yob,Bondarenko:2023fex}. We will return to its applications in Sec.~\ref{sec:case-studies}.

The semi-analytic approach is also well suited for estimating the sensitivity to FIP scatterings, which is the main signature in models of light dark matter. In this case, the differential decay probability should be replaced with the scattering probability 
\begin{equation}
\frac{dP_{\text{scatt}}}{d\theta dEdz} = n_{\text{detector}}\frac{d^{2}\sigma_{\text{scatt}}}{d\theta dE},
\end{equation}
where $n_{\text{detector}}$ is the number density of target particles inside the detector, and $d^{2}\sigma_{\text{scatt}}/d\theta dE$ is the differential cross-section for the scattering of FIPs off the target particles.

\subsection{Validation and limitations}
\label{sec:validation}
\begin{figure*}
    \centering
    \includegraphics[width=0.48\textwidth]{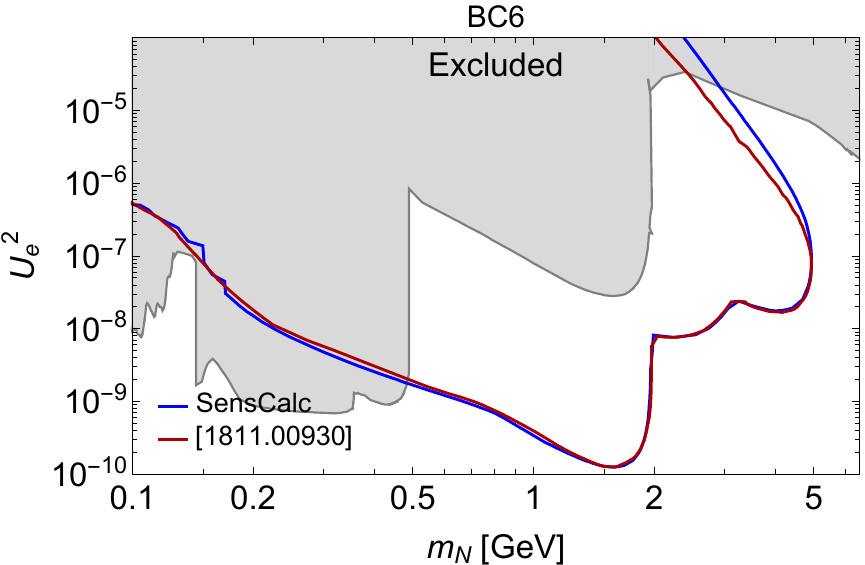}
    \hfill
    \includegraphics[width=0.48\textwidth]{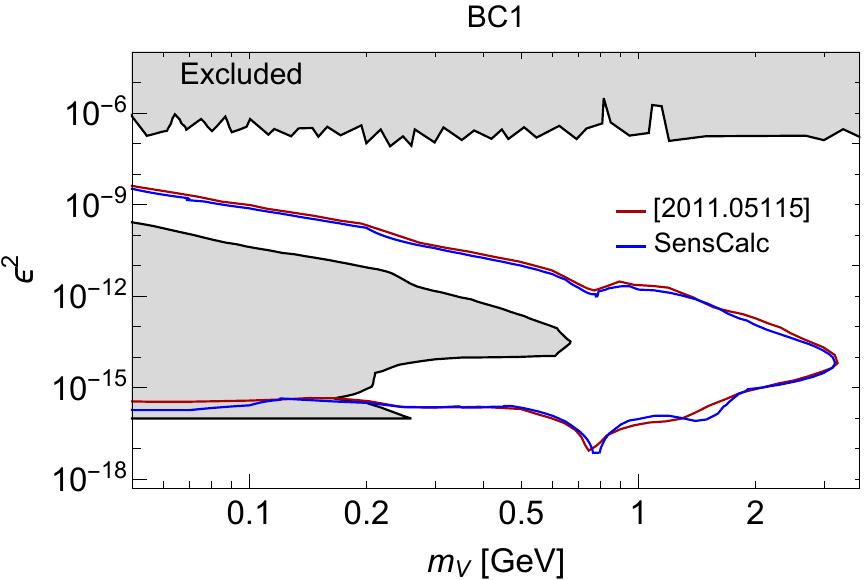}
    \caption{Comparison of the 90\% CL sensitivity of the SHiP experiment to heavy neutral leptons (\textbf{left panel}) and dark photons (\textbf{right panel}), obtained using Eq.~\eqref{eq:Nevents} within the framework of \texttt{SensCalc} and derived using the \texttt{FairShip} simulations~\cite{SHiP:2020vbd,SHiP:2018xqw}. The old ECN4 configuration of SHiP has been considered here.}
    \label{fig:validation}
\end{figure*}

The semi-analytic approach presented above has been used to estimate the sensitivities of various experiments at the SPS~\cite{Boiarska:2021yho}, LHC~\cite{Boyarsky:2022epg,Ovchynnikov:2022its}, and FCC-hh~\cite{Boyarsky:2022epg}. The considered experimental setups cover various options: on-axis and off-axis placements of the detector, different decay volume shapes, and different detector orientations relative to the beamline. These estimates, carried out using our semi-analytical method, have been found to agree well with the estimates available in the literature, including the simulations-based ones. In particular, Fig.~\ref{fig:validation} shows the comparison of the sensitivity of the SHiP experiment to heavy neutral leptons (HNLs) and dark photons obtained using Eq.~\eqref{eq:Nevents} with the sensitivity obtained by the SHiP collaboration using the \texttt{FairShip} simulation. In the case of dark photons, the slight differences in the sensitivity can be explained by the different elastic proton form factors used to describe the production probability. In the case of HNLs, the discrepancy at the upper bound follows from the monochromatic approximation of the HNL energy spectrum assumed when computing the sensitivity shown in the SHiP paper~\cite{SHiP:2018xqw}.

\begin{table*}
    \centering
    \begin{tabular}{|c|c|c|c|c|c|c|}
      \hline Experiment & SHADOWS & MATHUSLA@CMS   \\ \hline
      $(x,y, z)_{\text{min}}$, m  & (-1,0,14) & (0,60,68) \\ \hline
     Fid.\ dim, $\text{m}^{3}$   & $2.5\times 2.5\times 20$  & $100\times 25\times 100$ \\ \hline
      Det.\ dim., $\text{m}^{3}$  & $2.5\times 2.5\times 12$ & $100\times 5\times 100$ \\ \hline
      Detector plane & $xy$ & $xz$ \\ \hline
       \makecell{Requirement for \\ decay products}   & \makecell{Point to the end of detector \\ Oppositely charged, or neutral\\ No other cuts} & \makecell{Point to the end of detector \\ Oppositely charged \\ No other cuts} \\ \hline
       $B$ distribution & \cite{CERN-SHiP-NOTE-2015-009} & \cite{Kling:2021fwx} \\ \hline
        Scalar production      & \multicolumn{2}{c|}{Exclusive production,~\cite{Boiarska:2019jym}} \\ \hline
         Scalar decays & \multicolumn{2}{c|}{Following~\cite{Boiarska:2019jym}} \\ \hline
    \end{tabular}
    \caption{Description of the experimental setups and of the scalar phenomenology used to obtain the sensitivity shown in Fig.~\ref{fig:validation-off-axis}. The rows indicate, respectively, the closest distance from the collision point to the decay volume (the $z$ axis being along the beamline), the decay volume dimensions, the detector dimensions, the orientation of detector layers, the decay products acceptance criteria, the distribution of $B$ mesons used to calculate the flux of scalars, the scalar production branching ratios, and the description of the scalar lifetime and decays. The description of the experiments has been taken from Refs.~\cite{Alviggi:2839484} (SHADOWS) and~\cite{MATHUSLA:2022sze} (MATHUSLA@CMS). For the description of the scalar production, we followed the PBC recommendations~\cite{Beacham:2019nyx}.}
    \label{tab:comparison-with-FIPMC}
\end{table*}

\begin{figure*}
    \centering
    \includegraphics[width=0.48\textwidth]{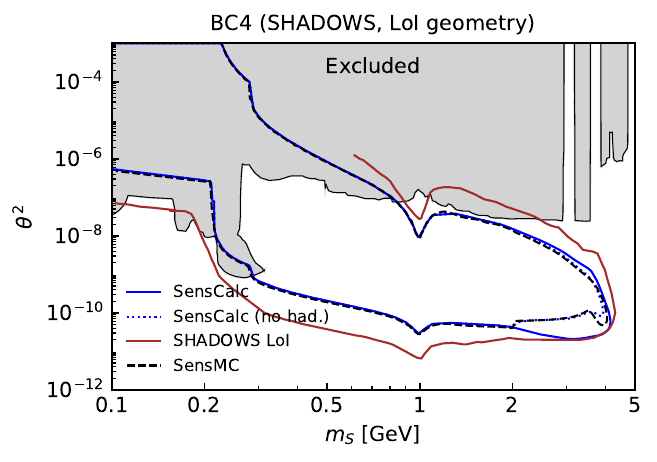}
    \hfill
    \includegraphics[width=0.48\textwidth]{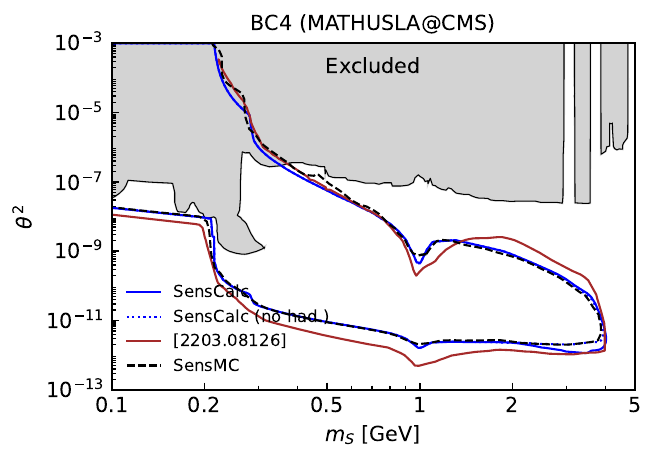}
    \caption{Comparison of the predictions of \texttt{SensCalc} (the blue lines) with the \texttt{SensMC} Monte-Carlo code used for validation (dashed black; described in App.~\ref{app:toy-mc}), for the sensitivity of the experiments located off-axis. SHADOWS (\textbf{left}, with the setup described in Ref.~\cite{Alviggi:2839484}) and MATHUSLA~\cite{MATHUSLA:2022sze} (\textbf{right}) are considered for the comparison. The description of the experiments has been taken from the collaboration papers. For SensCalc, we show two curves: the dashed line, for which the decays into partons are treated without the hadronization (this curve is to be compared with \texttt{SensMC}), and the solid line, for which hadronization is included. \texttt{SensMC} does not include the showering and hadronization of partonic decay products nor the dipole magnet's effect on the decay products' trajectories. Therefore, to compare \texttt{SensCalc} and \texttt{SensMC} under consistent assumptions, for SHADOWS, the effect of the dipole magnet in \texttt{SensCalc} has been turned off, and two sensitivities are shown: one where the hadronic decays are treated without hadronization (dashed line), and one where hadronization is included (solid line). All the characteristic quantities produced by the two approaches mostly agree within 20\% (see text for detail). Discrepancies are caused mostly by different treatments of scalar decay. The solid red lines show the sensitivities reported in the collaboration documents~\cite{Alviggi:2839484,MATHUSLA:2022sze} (see text for discussions).}
    \label{fig:validation-off-axis}
\end{figure*}

If the assumptions are well-controlled, the semi-analytic approach can agree very well with simulations. Fig.~\ref{fig:validation-off-axis} compares the sensitivity of SHADOWS and MATHUSLA to dark scalars as computed via Eq.~\eqref{eq:Nevents} and calculated independently by \texttt{SensMC}, a simple weight-based Monte-Carlo that we have implemented as described in Appendix~\ref{app:toy-mc} (see Table~\ref{tab:comparison-with-FIPMC} for the detailed description of the setup and of the scalar phenomenology used to compute the sensitivity). In these calculations, we did not impose any cuts on the decay products apart from the geometric requirement $\epsilon_{\text{decay}}^{\text{geom}}$; therefore, the sensitivities shown are optimistic. The agreement between the two approaches is within 10--20\% depending on the scalar mass; the discrepancies may be explained by numeric differences in the total number of produced scalars, the sampling of the distribution of $B$ mesons (for MATHUSLA), and slightly different treatments of the decay chain of the scalar.

As a further demonstration of the importance of having open-access sensitivity calculations with clear and controllable assumptions and inputs, we have also included in Fig.~\ref{fig:validation-off-axis} the sensitivities reported in the respective collaboration papers: the SHADOWS LoI~\cite{Alviggi:2839484}, and the MATHUSLA EoI~\cite{MATHUSLA:2022sze}. These sensitivities differ greatly from those we obtained for two main reasons. First, both collaborations use a different description of the scalar production based on the inclusive estimate of the decay of $B$ mesons. Namely, such a decay into a dark scalar is described as the decay of the $B$ meson's constituent $b$ quark. Second, the assumptions about the experimental setups used to compute the sensitivity differ from what the documents describe. In the case of SHADOWS, Ref.~\cite{Alviggi:2839484} used not the setup described within that same work (and summarized in Table~\ref{tab:comparison-with-FIPMC}), but a more optimistic setup located closer to the target and the beamline.\footnote{From private communications with the representatives of the SHADOWS and MATHUSLA collaborations.} In the case of MATHUSLA, the acceptance of the decay products was assumed to be~$1$ in Ref.~\cite{MATHUSLA:2022sze}, which may be too optimistic.\footnote{This is due to two reasons. First, the detector covers only the upper wall of the decay volume, which is parallel to the beamline, and not the other walls; this restricts the angular acceptance of the decay products. Second, FIPs decaying inside MATHUSLA have low energies. As a result, their decay products have a large angular spread.} These differences can significantly affect the reported sensitivity.

Finally, the predictions of our method agree with other publicly available packages --- \texttt{FORESEE} and \texttt{ALPINIST}, as will be discussed in more detail in Sec.~\ref{sec:comparison-with-codes}.\footnote{For other packages, see also~\cite{Harland-Lang:2019zur,Buonocore:2018xjk,deNiverville:2016rqh}.}

The simplicity of our semi-analytic method incurs some limitations. First, it cannot provide the full event record associated with each FIP decay or interaction, i.e.,\ the set of all initial, intermediate, and final-state particles, including their full kinematics. Instead, it averages over all events that pass the selection. Therefore, it does not allow studying the reconstruction of the FIP parameters, such as its mass, for which detailed event information is essential. Second, the approach assumes that the surrounding infrastructure does not influence the production of the FIPs. While this is often true in the case of FIPs produced at the collision point or close to it, the situation is different for non-prompt production, e.g., the production in decays of long-lived $K^{\pm}$ or $K^{0}_{L}$ mesons, from neutrino up-scatterings (the neutrino dipole portal~\cite{Ovchynnikov:2022rqj,Ballett:2019bgd}), or the conversion of photons into axion-like particles (ALPs) in the magnetic field at the LHC~\cite{Kling:2022ehv}.

\section{SensCalc}
\label{sec:senscalc}

\subsection{Description}
\label{sec:description}

\begin{table*}
    \centering
    \begin{tabular}{|c|c|c|}
    \hline Facility & List of experiments  \\ \hline
       SPS  & SHiP~\cite{Aberle:2839677,Ahdida:2867743}, NA62$_{\text{dump}}$~\cite{CortinaGil:2839661}, HIKE$_{\text{dump}}$~\cite{CortinaGil:2839661,Ahdida:2867743}, SHADOWS~\cite{Alviggi:2839484,Ahdida:2867743} \\ \hline
        Fermilab (dump)     & DUNE and DUNE-PRISM~\cite{DUNE:2021tad}, DarkQuest~\cite{Batell:2020vqn} \\ \hline
        LHC   & \makecell{FASER/FASER2/FASER$\nu$/FASER$\nu 2$~\cite{FASER:2019aik,FASER:2020gpr,Feng:2022inv}\\ SND@LHC/advSND~\cite{SHiP:2020sos,Feng:2022inv}\\  FACET~\cite{Cerci:2021nlb}, MATHUSLA~\cite{MATHUSLA:2022sze}, CODEX-b~\cite{Aielli:2019ivi} \\ ANUBIS in the shaft and ceiling configurations~\cite{Bauer:2019vqk} \\ LHCb}  \\ \hline FCC-hh & Analogs of the LHC-based experiments~\cite{Boyarsky:2022epg} \\ \hline
    \end{tabular}
    \caption{List of the experiments whose geometry is currently implemented in \texttt{SensCalc}, along with, for each experiment, a reference containing a description of the setup used.}
    \label{tab:implemented-experiments}
\end{table*}

\begin{figure*}
    \centering
    \includegraphics[width=0.8\textwidth]{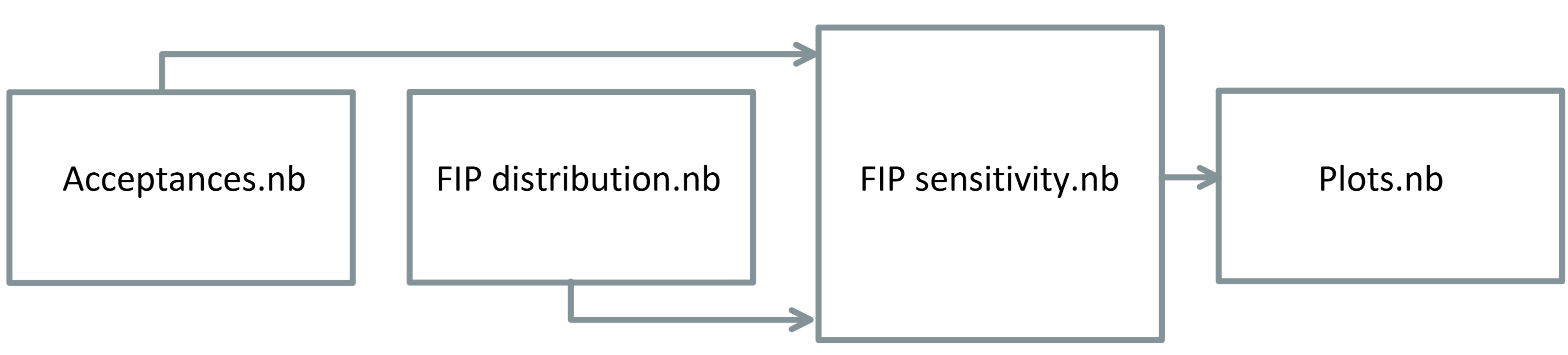}
    \caption{Sketch of the modular structure of \texttt{SensCalc}. The notebook \texttt{Acceptances.nb} produces the list of acceptances $\epsilon_{\text{az}}$ and $\epsilon_{\text{dec}}$ entering Eq.~\eqref{eq:Nevents} for the selected experiment. The notebook \texttt{FIP distribution.nb} computes the distribution of FIPs $f(m,\theta,E)$ at the facility housing the experiment. The notebook \texttt{FIP sensitivity.nb} uses as input the outputs of the two previous notebooks to calculate the tabulated number of events, and then calculates the sensitivity in the mass-coupling plane as a function of the remaining parameters such as the minimal number of events and any additional model-specific parameters. Finally, \texttt{Plots.nb} produces the sensitivity plots from the output of the previous notebook.}
    \label{fig:senscalc-modules}
\end{figure*}

The code \texttt{SensCalc} consists of a few \texttt{Mathematica} notebooks that compute the number of events for various FIPs (see Table~\ref{tab:implemented-models-production} for the list of the currently available models). Four notebooks have to be run sequentially: \texttt{Acceptances.nb}, \texttt{FIP distribution.nb}, \texttt{FIP sensitivity.nb}, and \texttt{Plots.nb}, see Fig.~\ref{fig:senscalc-modules}.

\begin{figure*}
    \centering
    \hfill
    \includegraphics[width=0.42\textwidth]{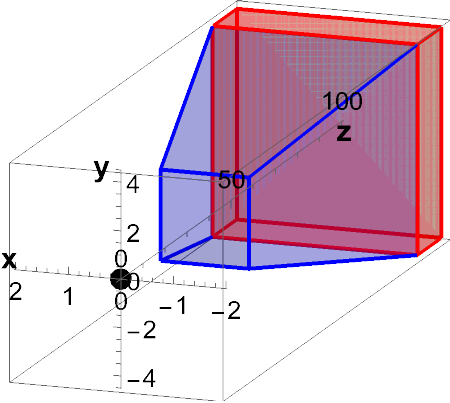}
    \hfill\hfill
    \includegraphics[width=0.42\textwidth]{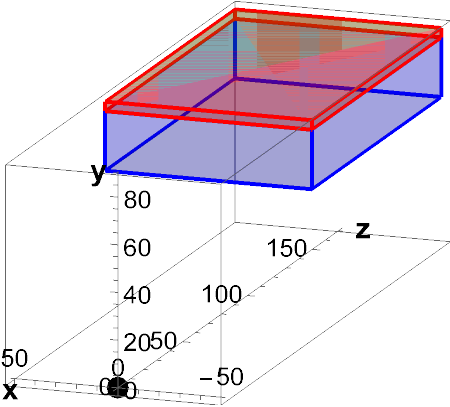}
    \hfill\hfill
    \caption{Visualizations of the geometries of the SHiP (\textbf{left}) and MATHUSLA (\textbf{right}) experiments, as implemented in \texttt{SensCalc} (in the notebook \texttt{Acceptance.nb}). The blue domain corresponds to the decay volume, while the red domain shows the detector. The descriptions of the two geometries have been taken from the SHiP LoI~\cite{Aberle:2839677} and Ref.~\cite{MATHUSLA:2022sze}.}
    \label{fig:first-module}
\end{figure*}

\paragraph{In the first notebook, \texttt{Acceptances.nb},} the user specifies the experimental setup --- the geometry and dimensions of the decay volume and detector, as well as some details about the detector, such as the presence of an ECAL and dipole magnet and their parameters, see Fig.~\ref{fig:first-module}. The list of the experiments currently implemented in \texttt{SensCalc} is provided in Table~\ref{tab:implemented-experiments}. The user can easily implement new experiments or modify one of the already implemented setups, which may be useful when optimizing an experiment. Some past experiments are also included: CHARM~\cite{CHARM:1983ayi} and BEBC~\cite{BEBCWA66:1986err} at the SPS. In this notebook, the user must also provide all the relevant quantities, such as the number of protons on target (or the integrated luminosity for LHC- and FCC-hh-based experiments), the target material, and the production cross-sections for secondary particles (mesons and $W/Z/H$-bosons). For the implemented experiments, these parameters are already listed in the notebook.

Once the setup is fixed, the notebook evaluates the angular coverage of the experiment and $\epsilon_{\text{dec}}$ for various FIPs. Concretely, it first defines the grid of the FIP masses $m$, FIP energies $E$, and its decay coordinates within the decay volume: the polar angle $\theta$, the longitudinal displacement from the target along the beam axis $z$, and the azimuthal angle $\phi$. 

Let us describe the procedure of the grid generation in detail. The $m$ grid depends on the underlying FIP model. It covers the lightest and heaviest FIP that may be produced at the given facility. The grid is not very dense to reduce the computation time. To improve the quality of the sampling under these conditions, we sample in a non-uniform way: the mass grid is distributed such that it allows recovering abrupt changes to the kinematics when new decay channels open. For instance, for dark scalars, where decays into heavy fermions $f\bar{f}$ start dominating already close to the kinematic threshold, the mass grid includes the points left and right from thresholds of main decays and several intermediate masses to study the interplay between different decay modes. The grid in $E$ is logarithmic, starting from the FIP mass and ending with the maximal energy available at the given facility. The $z$ and $\theta$ grids are within the boundaries covered by the decay volume of the experiment. To improve the accuracy of the calculations, the $\theta$ grid is denser in the domains where the FIP points to the end of the detector (and which would provide the main contribution to the decay products acceptance). For each $\theta,z$, the notebook randomly generates $N$ values of the azimuthal angle $\phi \in (-\pi,\pi)$, checks if the point $\{\theta,z,\phi\}$ is inside the decay volume, and calculates the azimuthal acceptance $\epsilon_{\text{az}}(\theta,z) = M/N$, where $M\leq N$ is the number of $\phi$ values for which the point is inside. Then, it produces the list of $\phi$ for which the FIP is inside the decay volume.

Having the grid $(m,E,\{\theta,z,\phi
\}_{\text{inside decay volume}})$, the notebook then simulates the FIP decays using the relevant decay channels and calculates the decay acceptance $\epsilon_{\text{dec}}(m,\theta,E,z)$ by averaging over these decays and $\phi$. The averaging over $\phi$ is already possible at this stage since the other quantities that determine the number of events~\eqref{eq:Nevents} do not depend on the azimuthal angle. Namely, the differential decay probability $dP_{\text{dec}}/dz$ only depends on $z$ and $\theta$, while the FIP distribution function is typically isotropic in~$\phi$.\footnote{A possible exception is when FIPs are produced non-promptly, i.e., in decays of long-lived particles such as charged kaons.}

The decay channels implemented for each FIP are listed in Table~\ref{tab:implemented-models-decay}. By default, all decay channels that have at least two particles that may be reconstructed at the given experiment (it depends on the presence of the calorimeters and other limitations) are included in the computation. However, users may select their own list of decay channels. For 3-body decays, the distribution of the decay products is generated by taking into account both the phase space and the squared matrix element of the process. If the FIP decay products are short-lived, the routine decays them until only metastable particles are left. By default, those are $\gamma,e,\mu,K^{0}_{L},\pi^{\pm},K^{\pm}$. Some representative decays approximate the decays of SM particles with many modes; for example, for $\tau$, this is a 3-body decay into one charged particle and two neutrinos. 

The total rate of the hadronic decays of heavy FIPs $m\gtrsim 1\text{ GeV}$ may be calculated using perturbative QCD as decays into $GG$, $u\bar{u}$, $d\bar{d}$, $c\bar{c}$, $b\bar{b}$, etc. However, this information is not enough to calculate the decay products acceptance; this is because partons experience showering and hadronization, resulting in final states with large multiplicities. To estimate $\epsilon_{\text{dec}}$, users may choose one of two options for the phase space computation. One possibility is to consider the ``spectator approach'' by treating the partons as stable particles with a mass equal to the mass of the lightest charged hadron containing the given quark/gluon. Another option is to perform showering and hadronization of these partons to get a bunch of hadrons. For instance, the decay $\text{FIP}\to GG$ is treated either as a decay into two particles having the quantum numbers of charged pions, or into a bunch of $\pi,K,\gamma$ particles resulting from the showering and hadronization of the gluon pair. The corresponding phase space is pre-calculated for several characteristic FIP masses using \texttt{pythia8}, and then the resulting $\epsilon_{\text{dec}}$ may be interpolated (see details in Appendix~\ref{app:dis-simulation}); this procedure is accurate enough for our purposes. The impact of the hadronization is illustrated by Fig.~\ref{fig:validation-off-axis}, where we show the sensitivities of the SHADOWS and MATHUSLA experiments to dark scalars obtained using these two options.

Let us now discuss the computation of $\epsilon_{\text{dec}}$ in more detail. The main acceptance criterion is the requirement that the trajectories of at least two decay products with zero total electric charge are within the acceptance of the detector until its final plane. Decays into pure neutral final states (i.e., photons or $K^{0}_{L}$) are also included if a calorimeter is present. If the detector or decay volume includes a magnetic spectrometer, the components of the charged particles' coordinates and momenta are shifted by a kick right after the magnet to approximate the effect of the magnetic field. In addition to this geometric requirement, $\epsilon_{\text{dec}}$ may include various kinematic cuts on the visible final state particles resulting from the FIP's decays. The currently implemented cuts include cuts on the energy, transverse momentum, transverse impact parameter, and, for neutral particles in the calorimeter, their spatial separation. By complete analogy, the user may impose further kinematic cuts. Although the cuts are applied at the Monte-Carlo truth level, i.e.,\ they are implemented without considering reconstruction effects such as the finite resolution of 4-momenta measurements, they can already give us some understanding of the effects of a realistic event reconstruction on the signal yield. Such reconstruction effects could, in principle, be approximated by, e.g.,\ applying some smearing to the kinematics variables of the decay products, according to the detector resolution. Note that the acceptance criterion includes partially reconstructible states, i.e.,\ the final states for which the FIP invariant mass cannot be reconstructed from the detected decay products.

The output of the first notebook is a table with the following columns: 
\begin{equation}
\{m,\theta,E,z,\epsilon_{\text{az}},\epsilon_{\text{dec}}\}
\end{equation}

\begin{table*}
    \centering
    \begin{tabular}{|c|c|c|c|c|c|}
    \hline Model & Ref. & Production channels \\ \hline
    BC1 &\cite{Ilten:2018crw,SHiP:2020vbd} & \makecell{Decays of $\pi,\eta,\eta'$, mixing with $\rho^{0}$\\ Proton bremsstrahlung, Drell-Yan process} \\
       \hline
        BC4, BC5 &\cite{Boiarska:2019jym}& \makecell{2-/3-body decays of $B$, decay $h\to SS$ \\ Proton bremsstrahlung} \\ \hline
       BC6--8 & ~\cite{Bondarenko:2018ptm}& \makecell{2-/3-body decays of $B,D,W$}  \\ \hline
         BC9 & \cite{Jerhot:2022chi,Dobrich:2019dxc} & \makecell{Coherent 
 production: Primakov process, $pZ$ scattering \\ Decays of $\pi^{0},\eta$} \\ \hline
          BC10 & \cite{DallaValleGarcia:2023xhh}  & \makecell{Decays of $B$, mixing with $\pi^{0}/\eta/\eta'$ \\ Deep-inelastic production} \\ \hline
           BC11 & \cite{Aloni:2018vki,Chakraborty:2021wda,Jerhot:2022chi} & \makecell{Decays of $B$, mixing with $\pi^{0}/\eta/\eta'$ \\ Deep-inelastic production} \\ \hline
           \makecell{$U(1)_{B-L}$\\ $U(1)_{B-3L_{\mu}}$ \\ $U(1)_{B-L_{e}-3L_{\mu}+L_{\tau}}$\\ $U(1)_{B-3L_{e}-L_{\mu}+L_{\tau}}$} & \cite{Tulin:2014tya,Ilten:2018crw} & \makecell{Decays of $\pi,\eta,\eta'$, mixing with $\omega$\\ Proton bremsstrahlung\\ Drell-Yan process} \\ \hline
    \end{tabular}
    \caption{FIP production channels in the various models implemented in \texttt{SensCalc}. The columns are the model name (for those which are PBC benchmark models~\cite{Beacham:2019nyx}, we provide their identifier), the reference used to describe the production channels and the list of the production channels implemented in \texttt{SensCalc}. The models are dark photons (BC1), dark scalars with Higgs mixing (BC4) and also with the quartic coupling (BC5), heavy neutral leptons with arbitrary mixing patterns (including the limiting cases of the single-flavor mixing with $\nu_{e}$, $\nu_{\mu}$, or $\nu_{\tau}$ (BC6--BC8)), ALPs coupling to photons (BC9), fermions (BC10) and gluons (BC11), and anomaly-free mediators coupled to lepton and baryon numbers: $U(1)_{B-L}$, $U(1)_{B-3L_{\mu}}$, $U(1)_{B-3L_{e}-L_{\mu}+L_{\tau}}$, and $U(1)_{B-L_{e}-3L_{\mu}+L_{\tau}}$. See also Appendix~\ref{app:couplings-definitions} for a more detailed description of the models.}
    \label{tab:implemented-models-production}
\end{table*}

\paragraph{The second notebook, \texttt{FIP distribution.nb},} computes the angle-energy distribution of the FIPs produced by various facilities and mechanisms. The list of implemented production channels and relevant references used to describe the production can be found in Table~\ref{tab:implemented-models-production}. Many production mechanisms require knowing the distributions of the parent particles at the given facility, such as mesons, heavy SM bosons, and photons --- including those produced in secondary interactions. We provide them as tabulated distributions in polar angle and energy, which we generate following the literature or just using available distributions from existing studies (see also Appendix~\ref{app:references} for a description of how we have generated the distributions of parent particles). Users may easily replace the included distributions with their own differential flux. With the distribution of parent particles at hand, we then derive the distribution of FIPs. If the FIPs are produced in decays, we compute their phase space in the rest frame of the parent particle and then boost it to the lab frame. In the case of 3-body decays, the phase space takes into account the matrix element of the process. For FIPs produced via elastic scattering, we adopt the differential cross-section of the process from existing studies and then convolve it with the distribution of the parent particles. Should the need arise, new production channels may be added by the user, following the above examples.

Such a derivation of the FIP distribution is not possible, however, in the case of FIPs that are produced inelastically in proton-proton collisions (such as via the Drell-Yan process for dark photons or deep-inelastic production of ALPs through the gluon coupling), which require an external simulation. In this case, we use \texttt{MadGraph5\_aMC@NLO} (v3.4.2) \cite{Alwall:2014hca} with a model implemented in \texttt{FeynRules}~\cite{Alloul:2013bka} and exported to the UFO format~\cite{Degrande:2011ua}. To account for showering and hadronization, the events simulated in \texttt{MadGraph} are further processed by \texttt{PYTHIA~8}~\cite{Sjostrand:2014zea}; see also Appendix~\ref{app:dis-simulation} for details. The UFO files and the tabulated FIP distributions are provided alongside \texttt{SensCalc}. 

\begin{figure*}
    \centering
    \includegraphics[width=0.47\textwidth]{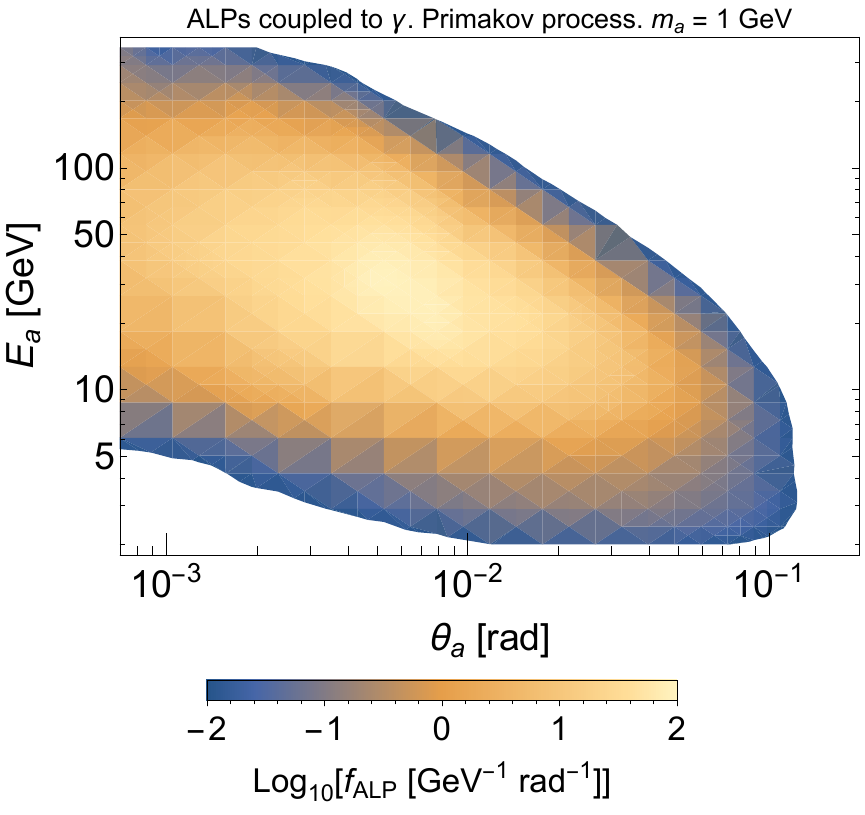}
    \hfill
    \includegraphics[width=0.47\textwidth]{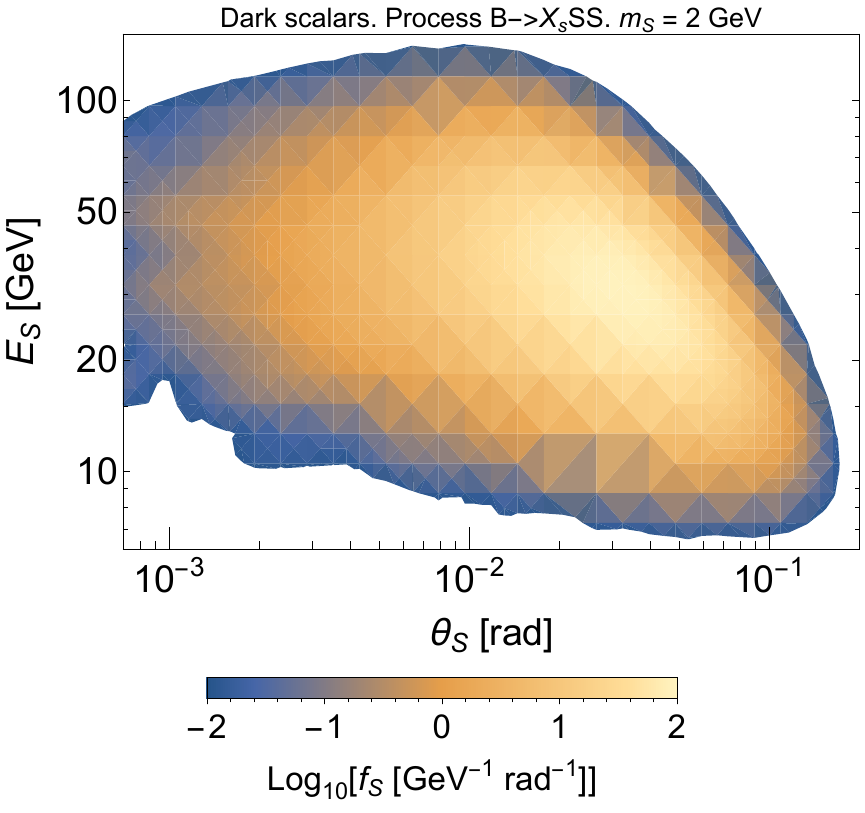}
    \caption{Examples of angle-energy distributions $f^{(i)}(\theta,E)$ for ALPs coupled to photons (\textbf{left}) and dark scalars with a non-zero quartic coupling (\textbf{right}), produced by the notebook \texttt{FIP distribution.nb}. The SPS facility with a molybdenum target is considered.}
    \label{fig:distributions-example}
\end{figure*}

The output of the second notebook is a tabulated distribution of the form 
\begin{equation}
\{m,\theta,E,f^{(i)}\},
\end{equation}
where the last column is the value of the FIP distribution function for the given $(m,\theta,E)$ and the production mechanism $i$. Some examples of computed distribution functions are shown in Fig.~\ref{fig:distributions-example}.

\begin{table*}
    \centering
    \begin{tabular}{|c|c|c|c|c|c|}
    \hline Model & Ref. & Decay channels (leptonic/$\gamma$) & Decay channels (hadr/semi-lept) \\ \hline
    BC1  &\cite{Ilten:2018crw,SHiP:2020vbd}& \makecell{$ee,\mu\mu,\tau\tau$ } & \makecell{$\pi\pi,3\pi,4\pi,KK, m\lesssim 2\text{ GeV}$\\ $q\bar{q}, m\gtrsim 2\text{ GeV}$}  \\
       \hline
        BC4, BC5 &\cite{Boiarska:2019jym,Winkler:2018qyg} & \makecell{$ee,\mu\mu,\tau\tau$} & \makecell{ $\pi\pi,KK,4\pi, m\lesssim 2 \text{ GeV}$ \\ $c\bar{c},s\bar{s},b\bar{b},GG, m\gtrsim 2 \text{ GeV}$} \\ \hline
       BC6-8 & \cite{Bondarenko:2018ptm} & \makecell{$3\nu, ll\nu$} & \makecell{ $\text{meson} + l/\nu, m\lesssim 1\text{ GeV}$\\ $\nu q\bar{q}, lq\bar{q}', m\gtrsim 1\text{ GeV}$}  \\ \hline
         BC9 & \cite{Jerhot:2022chi} & $\gamma \gamma$ &  \\ \hline
          BC10 & \cite{DallaValleGarcia:2023xhh}&   \makecell{$ee,\mu\mu,\tau\tau$} & \makecell{ $\gamma\pi\pi, \eta \pi \pi, 3\pi, 4\pi, m< 2.3\text{ GeV}$\\ $GG, m> 2.3\text{ GeV}$} \\ \hline
           BC11&  \cite{Jerhot:2022chi,Aloni:2018vki,ALP} & \makecell{$\gamma \gamma$} & \makecell{ $\gamma\pi\pi, \eta \pi \pi, 3\pi, 4\pi, m< 2.3\text{ GeV}$\\ $GG, m> 2.3\text{ GeV}$} \\ \hline \makecell{$U(1)_{B-L}$\\ $U(1)_{B-3L_{\mu}}$\\ \dots}&  \cite{Ilten:2018crw,Tulin:2014tya} & \makecell{$ee,\mu\mu,\tau\tau$}  & \makecell{$\pi^{0}\gamma,3\pi,KK, m< 1.6\text{ GeV}$\\ $q\bar{q}, m> 1.6\text{ GeV}$} \\ \hline
    \end{tabular}
    \caption{Decay channels of the FIPs implemented in \texttt{SensCalc}. From left to right: the model name (see the caption of Table~\ref{tab:implemented-models-production}), the reference used to describe the decays, and the decay channels into hadrons or a combination of hadrons and leptons. For dark scalars, we have included in their decay width into gluons the NLO correction from Ref.~\cite{Spira:1995rr}, which was previously missing in Ref.~\cite{Boiarska:2019jym}.}
    \label{tab:implemented-models-decay}
\end{table*}

Let us highlight an important point. Since the FIP distributions are determined mainly by the kinematics of the collisions, they can be considered identical for the different experiments housed at the same facility, assuming that the colliding particles are the same.\footnote{This is typically not the case for the non-prompt production of FIPs, which goes beyond the scope of the present discussion.} For collider experiments, we typically deal with proton-proton collisions, and this notebook only needs to be run once to obtain the distributions. In the case of beam dump experiments, some differences may arise due to different target/beam dump compositions. When the FIP is produced via the decays of secondaries, this only affects the overall scaling of the secondaries production cross-section, which depends on the atomic number $A$: $\sigma_{\text{prod,second}}\propto A^{0.29}$~\cite{Carvalho:2003pza}. Therefore, as in the collider case, the notebook only needs to be run once. If, however, the FIP is produced in scattering processes, then different targets may affect not only the normalization but also the shape of the distribution. To take this into account, we generate the fluxes for a few common types of targets.

\paragraph{The notebooks \texttt{<FIP> sensitivity.nb}} (with \texttt{<FIP>} replaced by the actual FIP) evaluate the sensitivity of the chosen experiment to the corresponding FIP. This is done via computing a tabulated number of events. First, the notebook imports the acceptance data computed by \texttt{Acceptances.nb}, the distributions produced by \texttt{FIP distribution.nb}, as well as the relevant quantities defining the FIP phenomenology, such as the production branching ratios, lifetimes, and branching ratios of the decays into visible states at the given experiment. It then maps them to a logarithmic scale and interpolates them to obtain the functions entering Eq.~\eqref{eq:Nevents}. 

Depending on the FIP, uncertainties in the description of its production and decay may significantly affect the event rate. This is the case, e.g., for dark scalars, where one may describe their production inclusively or exclusively; and for dark photons, for which the description of the proton bremsstrahlung channel depends on the maximal allowed $p_{T}$ and on the minimal energy allowed to be transferred to the dark photon. The user has the freedom to tune these parameters.

In addition, there may exist model-specific parameters that must be selected before performing the computation. For instance, in the case of HNLs, this is their nature (Dirac or Majorana) and mixing pattern $U_{e}^{2}:U_{\mu}^{2}:U_{\tau}^{2}$.

\begin{figure*}
    \centering
    \includegraphics[width=0.48\textwidth]{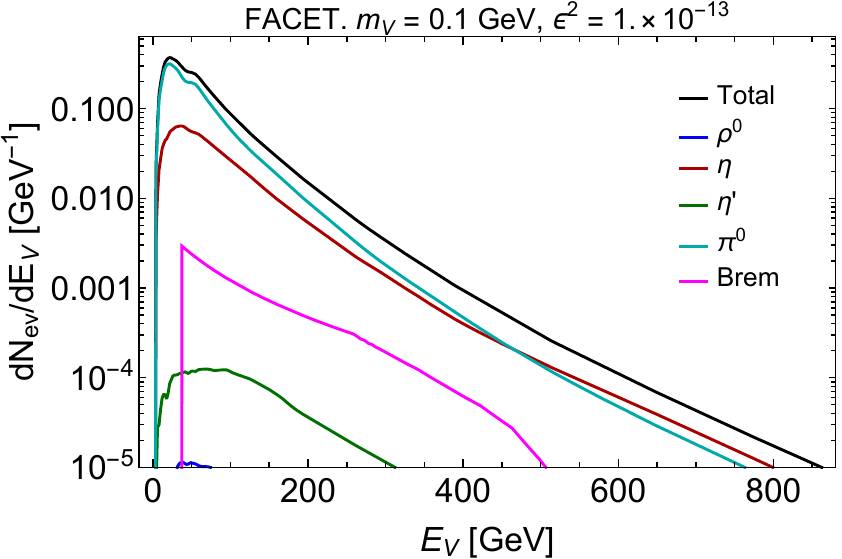}
    \hfill
    \includegraphics[width=0.48\textwidth]{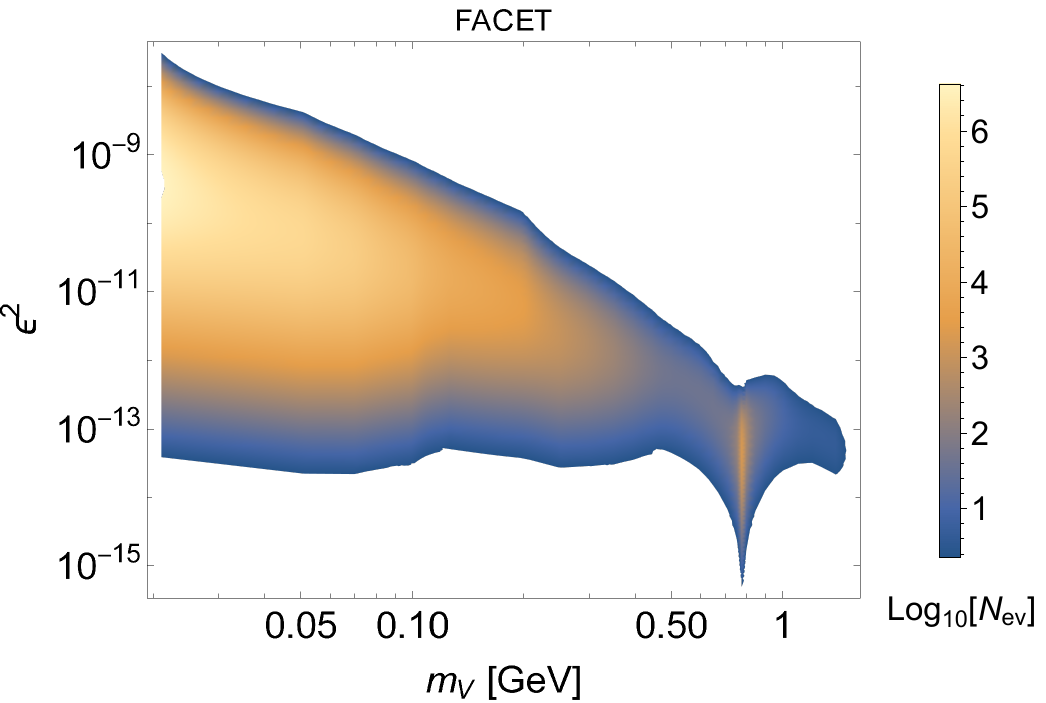}
    \caption{Examples of the output produced by the notebook \texttt{FIP sensitivity.nb}. \textbf{Left panel}: differential number of events with respect to the FIP's energy for various production channels. \textbf{Right panel}: the heatmap of the total number of events as a function of the FIP mass and coupling. As an example, dark photons at FACET are considered. No cuts on the decay products other than the geometric acceptance have been applied.}
    \label{fig:example-module-3}
\end{figure*}

During the computation, this notebook produces intermediate results that may be useful for the sensitivity analysis. This includes the differential number of events with respect to $\theta,E$, or $z$, as well as the number of events as a function of the mass and coupling (see Fig.~\ref{fig:example-module-3}). Last but not least, the notebook also outputs the overall acceptances $\epsilon$ (cf.\ Eq.~\eqref{eq:total-acceptance}) that may be used to quickly estimate the lower bound of the sensitivity and understand it qualitatively (see Sec.~\ref{sec:case-studies}).

Once the tabulated number of events has been produced, the notebook computes the sensitivities. To this end, the user needs to select the critical number of events determining the boundary of the sensitivity domain, as well as some model-specific parameters. For example, for dark scalars, one needs to specify the value of the branching ratio $\text{Br}(h\to SS)$, which is non-zero in the presence of the quartic coupling $\mathcal{L}\propto hSS$ (see Appendix~\ref{app:couplings-definitions} for details).
Because the critical number of events can be freely specified, the user can compute both ``exclusion'' sensitivity limits --- corresponding, e.g., to $2.3$ expected events at $90\%$~CL in the absence of background --- or ``discovery'' sensitivity limits by (externally) providing the critical $N_{\text{ev}}$ corresponding to the desired significance level and background expectation.

\begin{figure}[!h]
    \centering
    \includegraphics[width=\linewidth]{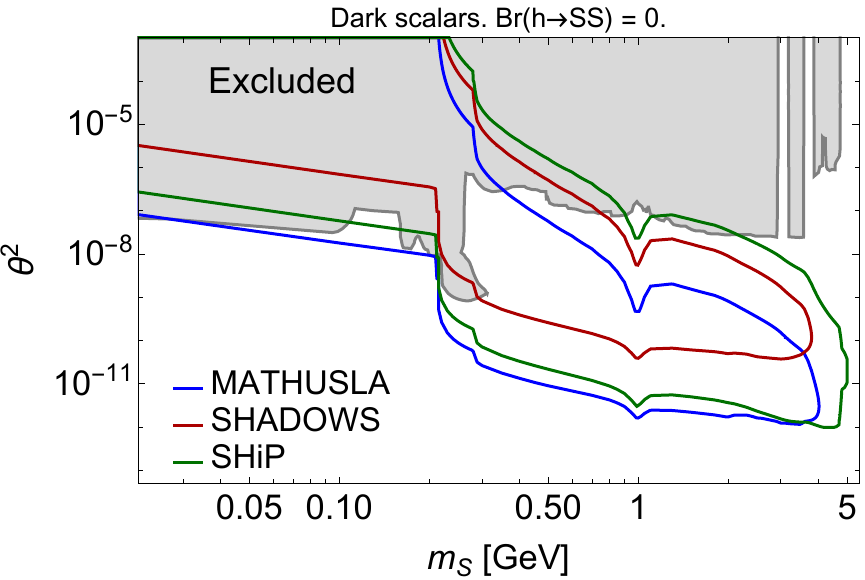}
    \caption{Example of a sensitivity plot produced by the notebook \texttt{Plots.nb}, for the model of dark scalars. The sensitivities of the SHiP, SHADOWS, and MATHUSLA experiments are reported. As for the description of the setups of the SHiP and SHADOWS experiments, we followed the latest document~\cite{Ahdida:2867743}. We assume that all the experiments operate in the background-free regime and define the sensitivity as $N_{\text{events}}>2.3$, corresponding to a 90\% CL limit.}
    \label{fig:example-module-4}
\end{figure}

\paragraph{Finally, the notebook \texttt{Plots.nb}} plots the sensitivities obtained in the previous notebook. It scans over the available sensitivity files, imports those needed by the user, and finally produces the figures (see e.g.\ Fig.~\ref{fig:example-module-4}).

\bigskip
The user interaction with the various notebooks, such as choosing the experiment, selecting the cuts, and the particular FIP model, is organized via dialog windows. This makes running the notebooks straightforward for FIPs and experiments that are already implemented.

To successfully run the notebooks, the user needs to install two dependencies: \texttt{FeynCalc}~\cite{Shtabovenko:2020gxv}, which is a \texttt{Mathematica} package for the symbolic evaluation of Feynman diagrams, and a \texttt{C} compiler that is recognized by \texttt{Mathematica}.

The performance of the code has been tested on various machines and operating systems. For instance, on a Windows laptop with 16~GB of RAM, 8 CPU cores, and \texttt{Mathematica 12.1}, the typical time required to compute the sensitivity from scratch is $\mathcal{O}(1\text{ hour})$ --- depending on the FIP type and on the mass-coupling grid density. This time is reduced if the FIP distribution has already been pre-generated.

\bigskip
\texttt{SensCalc} still offers significant potential for further improvement. Of particular interest would be the possibility to compute the sensitivity to additional FIP models, including those for which the main signature is scatterings with the detector material. Another well-motivated extension would be to support ALPs with an arbitrary coupling pattern. 


Finally, the implementations of the various experiments should be updated according to their latest specifications, which may differ from those listed in currently available documents. This may be done by contacting the representatives of the collaborations.

We are planning to add the above features in future code updates.

\subsection{Comparison with similar software packages}
\label{sec:comparison-with-codes}
At the moment of releasing \texttt{SensCalc}, there are two publicly available codes for computing the sensitivity of lifetime-frontier experiments to decaying FIPs: \texttt{FORESEE}~\cite{Kling:2021fwx} and \texttt{ALPINIST}~\cite{Jerhot:2022chi}.

\texttt{FORESEE} is a \texttt{Python}-based code developed to evaluate the sensitivities of the far-forward experiments at the LHC and FCC-hh. The currently implemented models of FIPs include dark scalars, dark photons, ALPs coupling to $W$ bosons, millicharged particles, and up-philic scalars. The package includes the tabulated distributions of various SM particles, including photons, mesons, and electroweak bosons. Apart from the tabulated number of events as a function of the FIP mass and coupling, it can additionally produce detailed event records in the \texttt{HepMC} format, which may then be passed to, e.g.,\ a detector simulation software. By default, \texttt{FORESEE} does not calculate the acceptance of the decay products; instead, it only requires the FIP to decay inside the decay volume, although the user may impose various cuts. It also does not hadronize partons.

\texttt{ALPINIST} computes the sensitivity of extracted-beam experiments --- including those at the SPS, Fermilab, and some past experiments --- to ALPs couplings to various SM particles. Its modules use~\texttt{Mathematica, ROOT}, and \texttt{Python}. The prominent feature of the code is that it can handle generic ALPs with simultaneous couplings to $W$ bosons, gluons, and the $U_{Y}(1)$ field. Unlike \texttt{FORESEE}, to obtain the tabulated number of events, the computation also incorporates the propagation of the decay products inside the detector, neglecting reconstruction effects such as the finite detector resolution. As a result, the computation time is much longer than for \texttt{FORESEE}. Only fully reconstructible final states are considered. The output of \texttt{ALPINIST} consists of data files with the mass-coupling dependence of the number of events for various production and decay modes. 

\begin{figure}
    \centering
    \includegraphics[width=\linewidth]{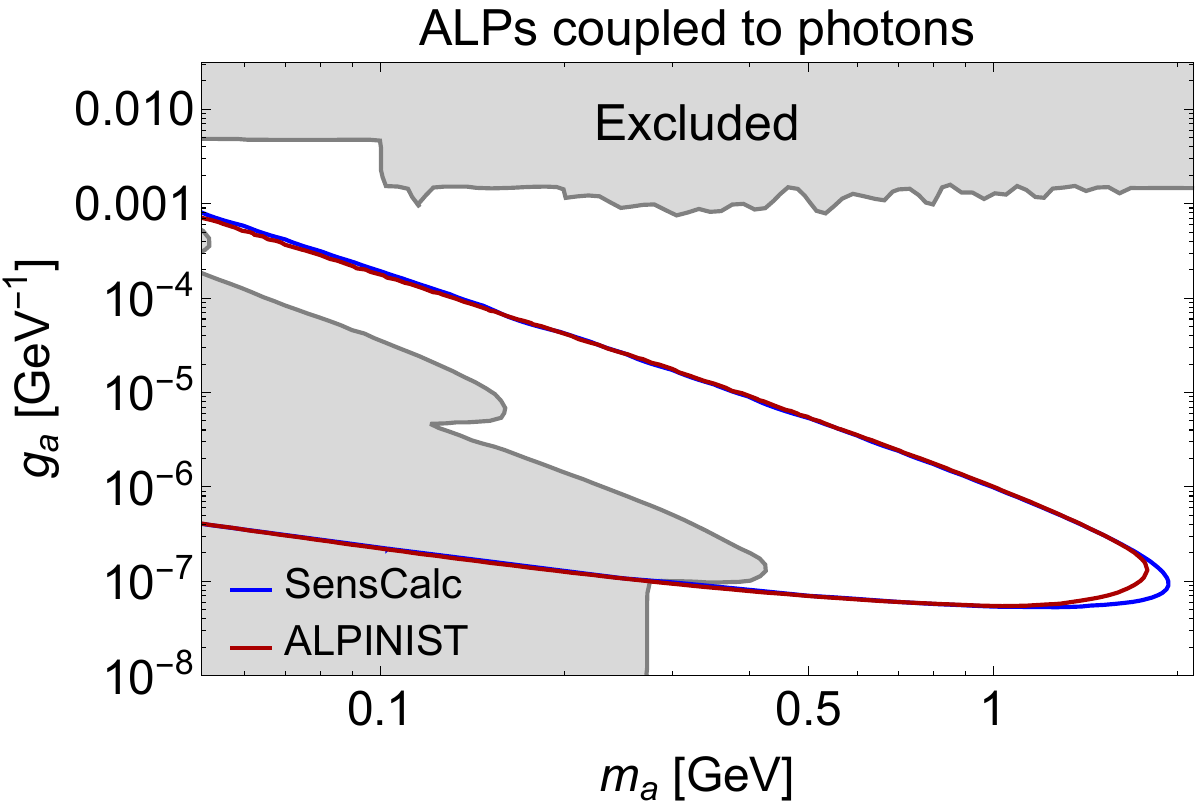}
    \caption{Comparison of the sensitivity of SHiP to ALPs coupling to photons as computed by \texttt{SencCalc} (blue line) and \texttt{ALPINIST}~\cite{Jerhot:2022chi} (red line). For the definition of the ALP coupling, see Appendix~\ref{app:couplings-definitions}. The SHiP configuration and the number of protons-on-target (that do not coincide with the configuration adopted by the SHiP collaboration) have been taken from the ALPINIST repository.}
    \label{fig:comparison-alpinist}
\end{figure}

The predictions of \texttt{SensCalc} agree well with the results of \texttt{ALPINIST} (see Fig.~\ref{fig:comparison-alpinist}) and FORESEE (the comparison between the semi-analytic approach and FORESEE is discussed in Ref.~\cite{Ovchynnikov:2022its}).

Unlike these two software packages, \texttt{SensCalc} is not restricted to a particular facility. In addition, among the implemented FIP models, it considers for the first time HNLs with arbitrary mixing patterns. The main limitation of \texttt{SensCalc} compared to \texttt{FORESEE} is that it cannot generate detailed event records, while compared to \texttt{ALPINIST}, it is that it does not (currently) consider generic ALPs and does not perform a detailed event reconstruction.

In addition, there have recently been a number of works related to the reinterpretation of experimental limits or sensitivities to FIPs, including Refs.~\cite{Tastet:2021vwp,Abada:2022wvh,Beltran:2023nli} and \cite[][Sec.\ 4.17]{Antel:2023hkf}.
Although related, these works are largely orthogonal (and thus complementary) to the present paper: while their aim is to reinterpret existing limits into new models that were not initially considered, our focus is to \emph{consistently} compute the sensitivity in the first place, for a restricted set of benchmark models.
This distinction is especially important when discrepancies exist in the assumptions used by different collaborations to report their sensitivities. Indeed, if those sensitivities were to be reinterpreted in a new model, the new limits would automatically inherit those same assumptions, allowing the discrepancy to propagate to the sensitivity plots of the new model.
By helping experiments report their limits under consistent assumptions, \texttt{SensCalc} can thus indirectly improve the consistency of reinterpreted limits as well.

\section{Case studies}
\label{sec:case-studies}

In this section, we demonstrate how \texttt{SensCalc} may be used by considering two examples: a qualitative understanding of the sensitivity reach for two particular experiments, and producing sensitivities for a particular FIP.

\subsection{Comparing two experiments: a detailed example}

\begin{figure*}
    \centering
    \includegraphics[width=0.45\textwidth]{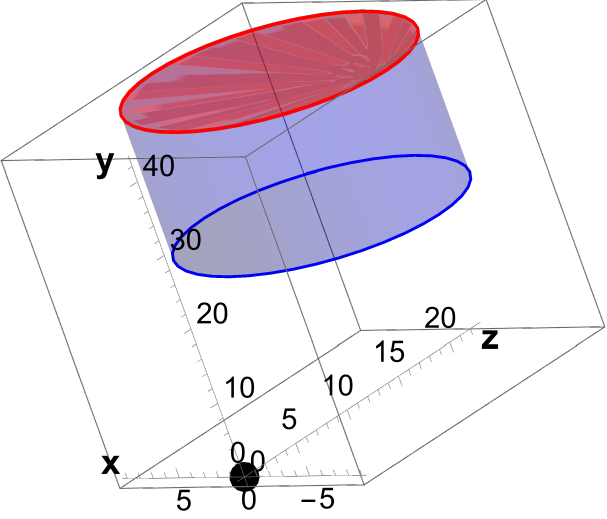}~\includegraphics[width=0.45\textwidth]{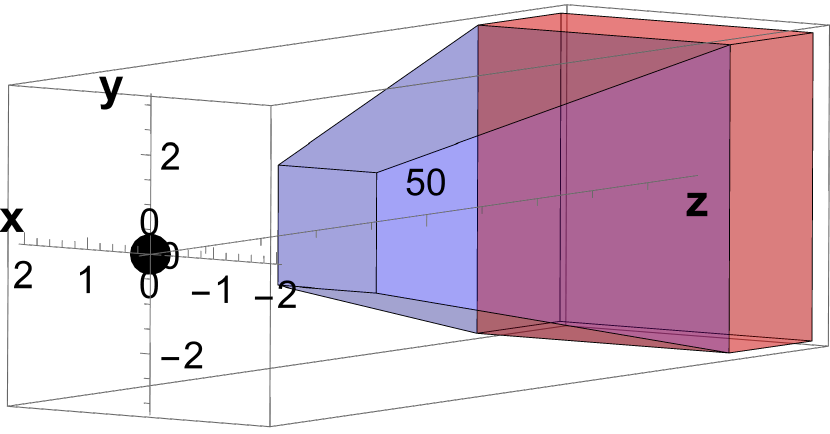}
    \caption{The first module of the ANUBIS experiment in the shaft configuration (left panel) and the SHiP experiment (right panel).}
    \label{fig:ship-anubis}
\end{figure*}

Consider, for example, two experiments --- SHiP and ANUBIS in the shaft configuration; see Fig.~\ref{fig:ship-anubis}.

They differ in all aspects. Namely, SHiP is to be located at the SPS, while ANUBIS --- at the LHC; it means completely different probabilities and angle-energy distribution of various production channels of FIPs. Next, SHiP is located on-axis, while ANUBIS-shaft is highly off-axis, with the detector plane being parallel to the beamline. Our goal is to understand qualitatively the impact of these differences on the behavior of the sensitivity for these two experiments in the regime of large lifetimes, for which the number of events may be represented as Eq.~\eqref{eq:number-of-events-large-lifetimes}. To analyze the number of events, we may start with the setting $\epsilon = 1$ (let us call the corresponding quantity $\mathcal{I}_{0} \equiv \sum_{i}N^{(i)}_{\text{prod}}$), and then sequentially include $f^{i}(\theta,E)\cdot \epsilon_{\text{az}}$ ($\mathcal{I}_{1}$), $dP_{\text{decay}}/dz$ ($\mathcal{I}_{2}$), and $\epsilon_{\text{dec}}$ ($\mathcal{I}_{3}$) in the integrand of Eq.~\eqref{eq:total-acceptance}. Their physical meaning will be the following: the total number of the FIPs produced at the given facility; the number of FIPs intersecting the decay volume; the number of FIPs decaying inside the decay volume; the number of FIP decays for which the decay products passed the decay acceptance.

\begin{figure}
    \centering
    \includegraphics[width=0.45\textwidth]{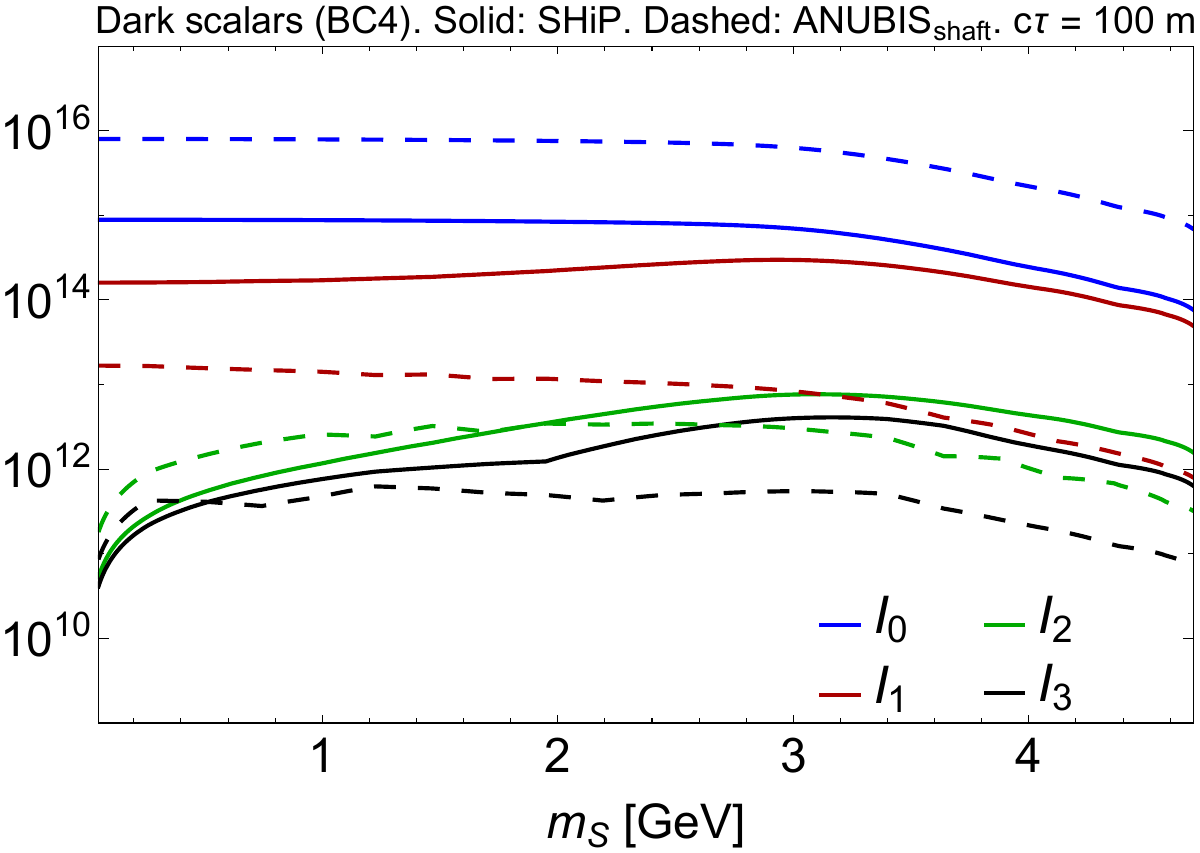}
    \caption{The behavior of the quantities $\mathcal{I}_{0-3}$ described in the text for the SHiP and ANUBIS-shaft configuration (all three modules are included). The overall normalization is arbitrary.}
    \label{fig:comparison_I0-I3}
\end{figure}

The comparison of the quantities $\mathcal{I}_{0}$--$\mathcal{I}_{3}$ for the model of dark scalars with the mixing coupling is shown in Fig.~\ref{fig:comparison_I0-I3}. There, for an apples-to-apples comparison, we assume no selection of the decay products except for the geometric requirement to point to the end of the detector.

The figure shows that the total number of produced FIPs is much larger at the LHC than at SPS. This is because the main production channel of the scalars is decays of $B$ mesons, whose production is more efficient for higher-energy proton collisions. Once the geometric placement of the decay volume and the detector are taken into account ($\mathcal{I}_{1}$), the situation changes. Namely, only a tiny fraction of $B$ mesons (and hence scalars) travels to the decay volume of ANUBIS, while for SHiP, the fraction is very significant, which results in a larger fraction of events at SHiP. Next, if one requires the FIP to decay ($\mathcal{I}_{2}$), the rates at SHiP and ANUBIS become similar; this is because the energy spectrum of the scalars at ANUBIS is much softer than at SHiP, which results in a larger decay probability (which scales as $\langle p_{S}^{-1}\rangle$). Finally, when adding the decay products acceptance requirement ($\mathcal{I}_{3}$), the number of events at ANUBIS decreases significantly compared to SHiP in the domain of large masses, which is explained by a larger angular spread of the decay products (making it more difficult for them to reach the detector) and the absence of a calorimeter at ANUBIS (so only charged decay products can be registered).

The situation may change if further acceptance requirements are added (e.g.\ to diminish the backgrounds), such as a minimal energy cut, to which ANUBIS-shaft is highly sensitive. However, this question will be the subject of another paper.

\subsection{ALPs coupled to fermions}

\begin{figure*}
    \centering
    \includegraphics[width=0.45\textwidth]{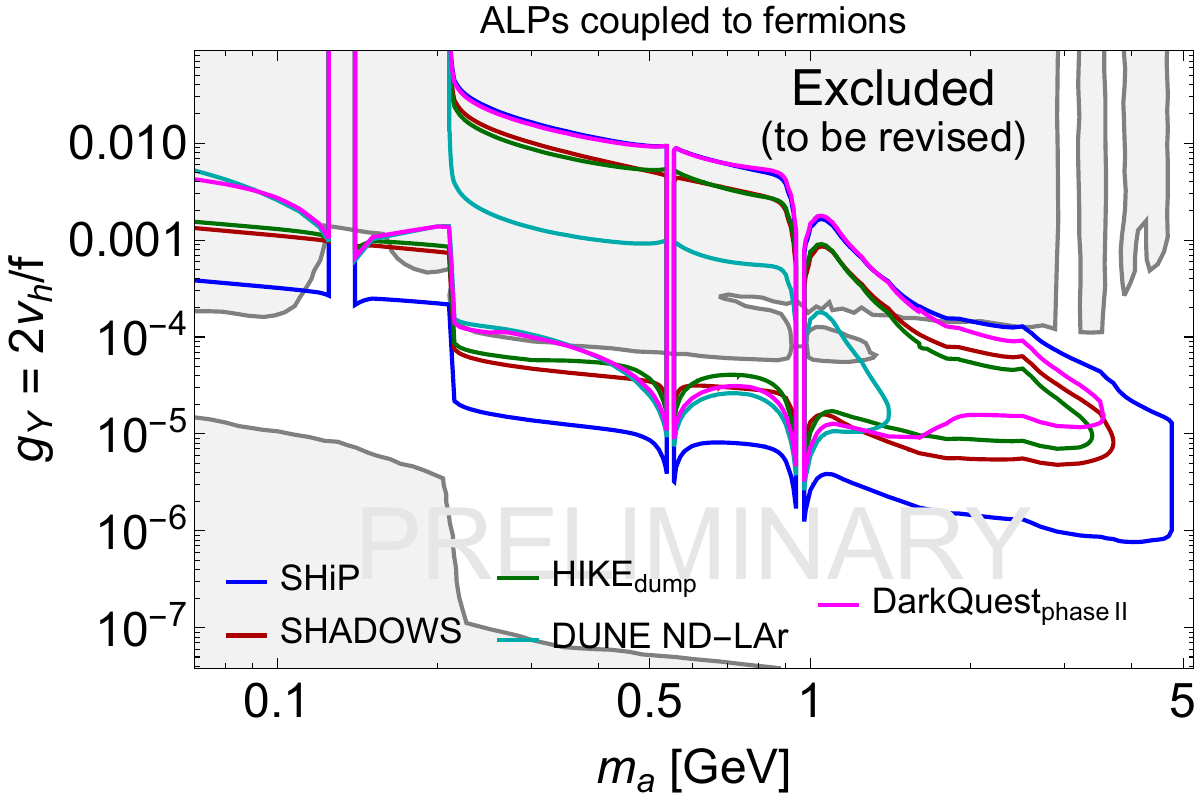}~\includegraphics[width=0.45\textwidth]{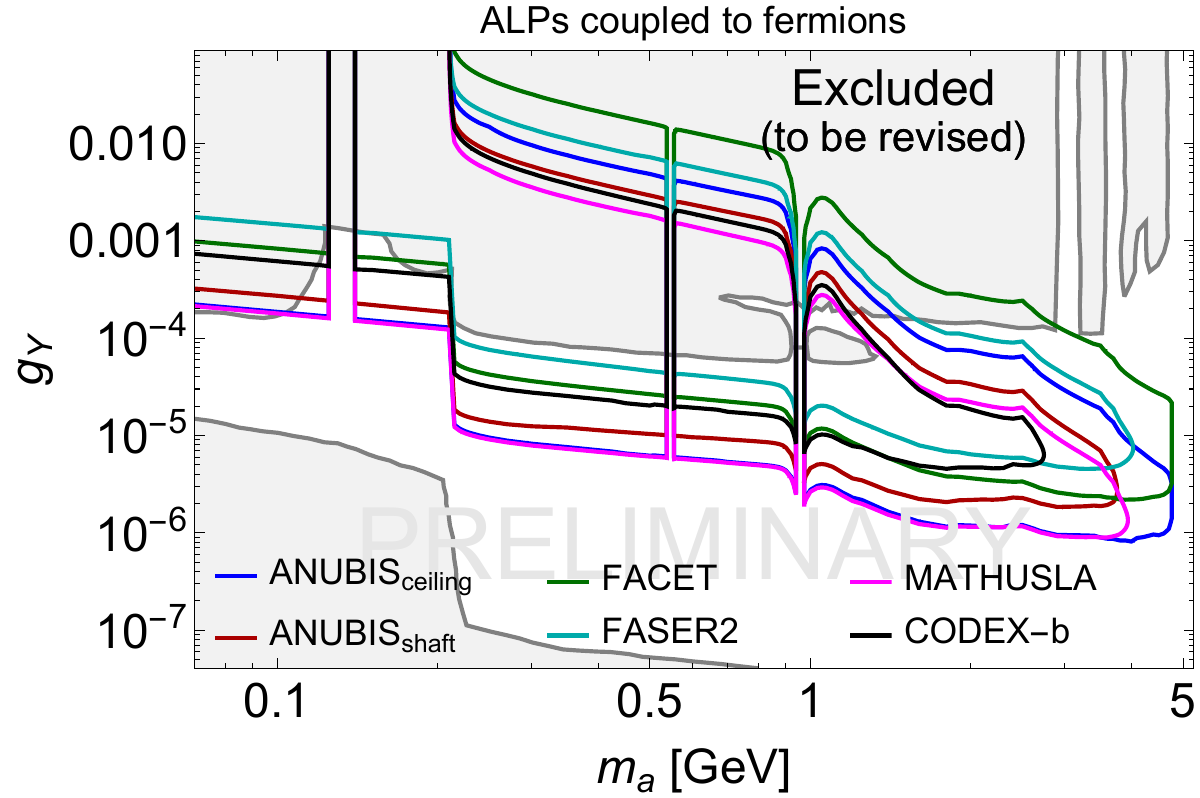}\\ \includegraphics[width=0.45\textwidth]{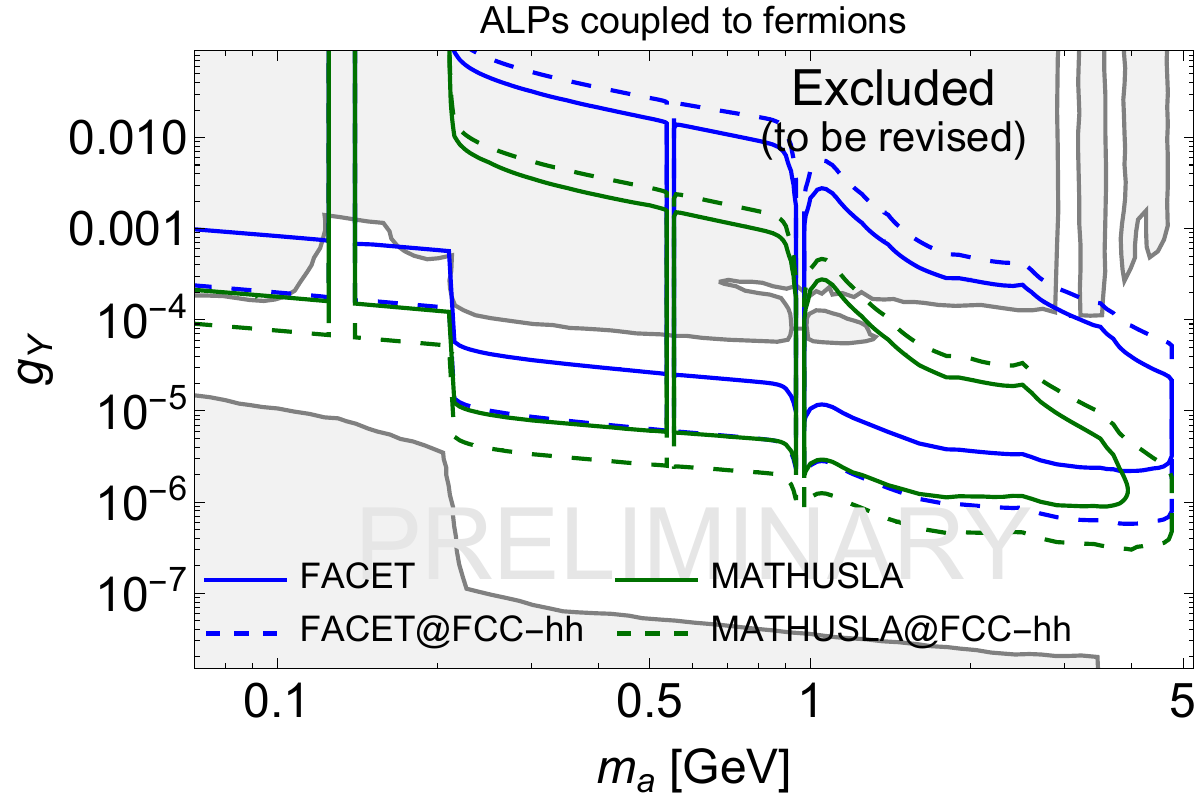}
    \caption{Sensitivities of the beam dump experiments (top left panel), LHC-based experiments (top right panel), and FCC-hh-based experiments (bottom panel) to the ALPs coupled to fermions considering the revised ALP phenomenology from~\cite{DallaValleGarcia:2023xhh}. For all the experiments, we used simplified assumptions of the absence of kinematic selection of the decay products other than geometric, absence of background, and considered 90\%CL sensitivity, corresponding to $N_{\text{events}}>2.3$.}
    \label{fig:sensitivities-alps}
\end{figure*}

As another example of \texttt{SensCalc} application, consider the model of axion-like particles coupled to fermions. The widely adopted ALP phenomenology description~\cite{Beacham:2019nyx} suffers from many issues; those include the absence of hadronic decays in the total decay width, which overestimates it by orders of magnitude for the ALPs with mass $m_{a}\gtrsim 1\text{ GeV}$, and the important production channels, such as $B$ decays into the ALP and heavy kaon resonances $K_{1},K_{2}^{*},K_{0}^{*}$, and the mixing with light neutral mesons $\pi^{0},\eta,\eta'$. The details are provided in~\cite{DallaValleGarcia:2023xhh}.

This means that the sensitivities of all the experiments to these ALPs have to be revised, which is a target subject of \texttt{SensCalc}. In Fig.~\ref{fig:sensitivities-alps}, we present the sensitivities of beam dumps, LHC-based, and some FCC-hh-based experiments to the ALPs. For the clarity of the comparison, we assume no background, the kinematic selection of the decay products other than the geometric requirement to pass to the end of the detector, and calculated 90\%CL sensitivities.

\section{Conclusion}
\label{sec:conclusions}
Feebly interacting particles (FIPs) are present in a broad class of new-physics scenarios that attempt to resolve the known problems of the Standard Model. Their search at various facilities and experiments collectively forms the lifetime frontier of particle physics. During the last decade, many lifetime-frontier experiments have been proposed, that differ in the housing facility, geometric location, and detector technology. With a few exceptions, most of these experiments are not approved yet, and their design is not finalized. Their sensitivities to FIPs are computed by the collaborations themselves, using internal tools which are not publicly accessible. This makes it difficult to control the inputs to the computations, such as the model of the production and decay. It is therefore crucial to have a publicly available tool for computing the sensitivity of those experiments to various FIPs in a uniform, fast and well-controlled way.

The present paper addresses this issue by presenting \texttt{SensCalc} --- a \texttt{Mathematica}-based code for evaluating the sensitivity of various experiments to decaying, long-lived FIPs, based on a semi-analytic approach developed in a number of previous studies (see Sec.~\ref{sec:method}) and cross-checked against various state-of-the-art packages (see Sec.~\ref{sec:validation}). 

\texttt{SensCalc} already supports a broad range of models and experiments (see Sec.~\ref{sec:description}). Models currently implemented include dark photons, dark scalars, heavy neutral leptons with various mixing patterns, axion-like particles coupled to different SM particles, and mediators coupled to anomaly-free combinations of the lepton and baryon currents. Numerous experiments have been implemented, located at any of the following facilities: the SPS, Fermilab (dump), LHC, and FCC-hh. The code is designed to be easily extended and could, in principle, support models of FIPs for which the main signature is scattering, as well as different facilities such as beam dumps with lepton beams colliding with the target. The user retains full control over every aspect of the sensitivity calculation, from the geometry of the experiment and the distribution of the FIP's parent particles to the branching ratios of the FIP production/decay modes and the requirements on the decay products. Besides contributing to the transparency and trustworthiness of the results, this also allows users to easily modify the underlying assumptions as needed, or to add their own models and experiments to \texttt{SensCalc}.

By publicly providing a transparent, semi-analytic method to consistently compute the expected signal at various lifetime-frontier experiments, \texttt{SensCalc} can help address the discrepancies that currently exist in the literature between the descriptions of FIPs and acceptances employed by different collaborations.
This is a timely and necessary contribution to the field of FIP searches, as many experiments are currently undergoing active development and optimization, while funding bodies and hosting facilities must decide which projects to prioritize.
\texttt{SensCalc} can help with the former by providing fast (re-)calculation of the expected signal as the experiment's design evolves, and with the latter by ensuring a fair and consistent comparison of the expected signals between the proposed experiments, with well-controlled assumptions thanks to a uniform and well-validated implementation of the official PBC benchmarks. This could be particularly relevant in the context of the ECN3 hall upgrade at the CERN SPS, in which a number of experiments are currently being considered for inclusion, namely HIKE, SHiP, and SHADOWS.

\begin{acknowledgments}
We thank Alexey Boyarsky and Oleg Ruchayskiy, who supervised the authors on the present topic in the past and helped develop the foundations of the approach described in this paper. We thank Felix Kahlhoefer and Jan Jerhot for helpful discussions on the phenomenology of ALPs and the \texttt{ALPINIST} code, and Felix Kling for discussions on \texttt{FORESEE}. We also thank Thomas Schwetz, Nashwan Sabti, Vsevolod Syvolap, Felix Kahlhoefer, and Inar Timiryasov for reading the manuscript at different stages of its writing. Finally, we thank the users of the \url{https://mathematica.stackexchange.com/} website, who greatly helped us optimize some elements of the code. MO received support from the European Union's Horizon 2020 research and innovation program under the Marie Sklodowska-Curie grant agreement No. 860881-HIDDeN. OM is supported by the NWO Physics Vrij Programme “The Hidden Universe of Weakly Interacting Particles” with project number 680.92.18.03 (NWO Vrije Programma), which is (partly) financed by the Dutch Research Council (NWO). KB is partly funded by the INFN PD51 INDARK grant.
JLT acknowledges partial financial support by the Spanish Research Agency (Agencia Estatal de Investigaci\'on) through the grant IFT Centro de Excelencia Severo Ochoa No CEX2020-001007-S, by the grant PID2019-108892RB-I00 funded by MCIN/AEI/ 10.13039/501100011033, by the European Union's Horizon 2020 research and innovation programme under the Marie Sk\l odowska-Curie grant agreement No 860881-HIDDeN, and by the grant Juan de la Cierva FJC2021-047666-I funded by MCIN/AEI/10.13039/501100011033 and by the European Union ``NextGenerationEU''/PRTR.
\end{acknowledgments}

\subsection*{Conflict of Interest Statement}

The authors of the present manuscript are also members of the SHiP collaboration, which represents one of the experimental proposals currently competing for funding and access to facilities, notably as part of the ongoing Physics Beyond Colliders study and in the context of the upcoming upgrade of the ECN3 hall at CERN. The present manuscript solely reflects the authors' views, and not those of the SHiP collaboration.

\appendix

\section{Uncertainties in the description of FIPs}
\label{app:couplings-definitions}

\subsection{Discrepancies in the literature}
The description of the FIP production and decay, and sometimes even the definition of the FIP couplings, may vary among the sensitivity estimates performed by the different collaborations. One example is the dark scalar~$S$. Following the PBC report~\cite{Beacham:2019nyx}, the SHiP collaboration uses the exclusive description of the production of $S$, while other collaborations adopt instead the inclusive description (see the discussion in Ref.~\cite{Boiarska:2019jcw}). In the domain $m_{S}\gtrsim 2-3\text{ GeV}$, where the inclusive approach breaks down, the difference in the number of produced scalars between these two descriptions may be a factor of 20 or more. Another problem arises from the theoretical uncertainty on the hadronic decay width, which may be as large as a factor of 100~\cite{Winkler:2018qyg,Monin:2018lee} (see also a recent discussion in Ref.~\cite{Gorbunov:2023lga}). While SHiP and SHADOWS assume the decay width computed in Ref.~\cite{Winkler:2018qyg}, the FASER collaboration~\cite{Feng:2017vli} uses the decay width from Ref.~\cite{Bezrukov:2009yw}. Depending on the calculation used, the sensitivity may therefore differ significantly.

Another example is with ALPs $a$ coupling to gluons. The PBC report defines an interaction of the form $\mathcal{L} \propto a g_{a}G^{\mu\nu,a}\tilde{G}_{\mu\nu}^{a}$, where $G^{\mu\nu}$ is the gluon field strength and $g_{a}$ is a fixed dimensionful coupling. Theoretical works often~\cite{Aloni:2018vki,Chakraborty:2021wda} adopt a different definition, $\mathcal{L} \propto a g_{s}^{2} g_{a}G^{\mu\nu,a}\tilde{G}_{\mu\nu}^{a}$, where $g_{s} = g_{s}(m_{a})$ is the QCD coupling. The latter definition is used by \texttt{ALPINIST}~\cite{Jerhot:2022chi} for computing the sensitivity of beam dump experiments to ALPs (and their results are used by the SHiP, HIKE, and SHADOWS collaborations in Ref.~\cite{Antel:2023hkf}). Furthermore, while some collaborations~\cite{Aielli:2019ivi} include the production of ALPs through gluon fusion, others do not (this is the case in particular of \texttt{ALPINIST}~\cite{Jerhot:2022chi}).

Another problem arises with ALPs that couple to fermions. The PBC~\cite{Beacham:2019nyx} recommends including only the decays into leptons in the total width --- even though it may be dominated by hadronic decays in the mass range $m_{a}\gtrsim 2m_{\pi}$ --- while some collaborations also include hadronic channels~\cite{Aielli:2019ivi}.

Such mismatches between the assumptions used to compute different sensitivities are particularly problematic when said sensitivities are shown in the same plot --- such as, e.g., in the FIPs~2022 proceedings~\cite{Antel:2023hkf} --- without emphasizing that the underlying assumptions differ.

\subsection{Definition of the FIP couplings used in \texttt{SensCalc}}

The effective Lagrangians of the models implemented in \texttt{SensCalc} are:
\begin{itemize}
    \item[--] \textbf{BC1} (dark photons):
    \begin{equation}
    \mathcal{L}_{\text{int}} = -\epsilon e V_{\mu}J^{\mu}_{\text{EM}}
    \end{equation}
     where $V_{\mu}$ is the dark photon field, $J_{\text{EM}}^{\mu}$ is the EM current, and $e = \sqrt{4\pi\alpha_{\text{EM}}}$ is the EM coupling.
    \item[--] \textbf{BC4} and \textbf{BC5} (dark scalars): 
    \begin{equation}
        \mathcal{L}_{\text{eff}} \supset m_{h}^{2}\theta hS + \frac{\alpha}{2}hS^{2}, 
    \end{equation}
    where $\theta$ is the mixing angle and $\alpha$ is the quartic coupling. By default, the sensitivity is evaluated assuming a constant branching ratio $\text{Br}(h\to SS) \propto \alpha^{2}$.
    \item[--] \textbf{BC6, BC7, BC8} (HNLs):
    \begin{multline}
       \hspace{1.5em}\mathcal{L}_{\text{int}} = \sum_{\alpha = e,\mu,\tau} U_{\alpha}\bar{N}\biggl(\frac{g}{\sqrt{2}} \gamma^{\mu}P_{L}l_{\alpha} W_{\mu} + \\ \frac{g}{2\cos(\theta_{W})}\gamma^{\mu}P_{L}\nu_{\alpha} Z_{\mu} \biggr)+\text{h.c.},
    \end{multline}
    where $N$ is the HNL, $U_{\alpha}$ the mixing angle, $g$ the weak coupling, and $l_{\alpha},\nu_{\alpha},W,Z$ the SM fields. The HNL may be either a Dirac or a Majorana particle.
    \item[--] \textbf{BC9} (ALPs coupling to photons):
    \begin{equation}
        \mathcal{L}_{\text{int}} = \frac{g_{a}}{4}a F_{\mu\nu}\tilde{F}^{\mu\nu},
    \end{equation}
    where $a$ is the ALP field, $g_{a}$ is a dimensionful coupling, and $F_{\mu\nu}, \tilde{F}_{\mu\nu} = \frac{1}{2}\epsilon_{\mu\nu\alpha\beta}F^{\alpha \beta}$ are the EM field strength and its dual.
    \item[--] \textbf{BC10} (ALPs coupling to fermions):
    \begin{equation}
        \mathcal{L}_{\text{int}} = \frac{g_{Y}}{2v_{H}}(\partial_{\mu}a)\sum_{\alpha}\bar{f}\gamma^{\mu}\gamma_{5}f,
    \end{equation}
    where $g_{Y}$ is a dimensionless coupling, $v_{H}\approx 246\text{ GeV}$ is the Higgs VEV, and $f$ are SM fermions.
    \item[--] \textbf{BC11} (ALPs coupling to gluons):
    \begin{equation}
        \mathcal{L}_{\text{int}} = g_{a}\frac{\alpha_{s}}{4\pi}a G_{\mu\nu}^{a}\tilde{G}^{\mu\nu,a},
    \end{equation}
    where $g_{s}$ is the strong coupling constant, $a$ is the ALP field, $g_{a}$ is a dimensionful constant, $G_{\mu\nu}^{a}$ is the gluon field strength, and $\tilde{G}_{\mu\nu}^{a} = \frac{1}{2}\epsilon_{\mu\nu\alpha\beta}G^{\alpha\beta,a}$ is its dual field strength. Everywhere except for the production of ALPs from DIS, we follow the definition of $g_{s}$ from Ref.~\cite{Jerhot:2022chi}. In the DIS case, we employ the running of $g_{s}$ associated with the default PDF set in \texttt{MadGraph}.
    \item[--] Mediators coupled to the anomaly-free combinations of the baryon and lepton numbers:
    \begin{equation}
        \hspace{0.5em}\mathcal{L}_{\text{int}} =\sqrt{4\pi\alpha_{B}}\sum_{f}V^{\mu}Q_{f}\bar{f}\gamma_{\mu}(c_{f}-a_{f}\gamma_{5})f,
    \end{equation}
    where $V_{\mu}$ is the mediator, $\alpha_{B}$ is the coupling constant, and $Q_{f}$ are charges corresponding to the given group. For instance, for $B-L$ group, they are $Q_{e,\mu,\tau,\nu} = -1$ for leptons and $Q_{u,c,t,d,s,b} = 1/3$ for quarks. For $B-3L_{\mu}$, the lepton charges are $Q_{\mu}=Q_{\nu_{\mu}} = 3$, and $Q_{e,\tau,\nu_{e},\nu_{\tau}} = 0$. The coefficients are $c_{f} = 1, a_{f} =0$ for all fermions except for neutrinos. For the latter, $c_{f} = a_{f} = 1/2$. The implemented models are $B-L$, $B-3L_{\mu}$, $B-3L_{e}-L_{\mu}+L_{\tau}$, and $B-L_{e}-3L_{\mu}+L_{\tau}$.
\end{itemize}

\section{Inputs used for generating the signal yield}
\label{app:references}

\subsection{Deep-inelastic scattering production and decays into light partons}
\label{app:dis-simulation}

There are two types of processes with FIPs for which it is not possible to properly calculate the phase space in \texttt{Mathematica}: deep inelastic scattering (DIS) production (such as gluon fusion) and hadronic decays at scales $m_{\text{FIP}}\gg \Lambda_{\text{QCD}}$. At the hard level, these processes are just parton fusion into FIPs and decays into light partons. The resulting kinematics and final state multiplicity depend strongly on the subsequent showering and hadronization.

To calculate the cross sections and the FIP/decay products distributions for these processes properly, we implement the relevant interactions of the FIPs with quarks and gluons in \texttt{MadGraph5\_aMC@NLO} using \texttt{FeynRules}. Then, we simulate the production and decay processes in \texttt{MadGraph}, interfaced with \texttt{PYTHIA~8} for showering and hadronization. 

The hard processes that we simulate are the leading-order and next-to-leading-order processes for quark and gluon fusion:
\begin{equation}
\begin{aligned}
    q+\bar{q}\to V&, \quad  q+\bar{q}\to V + j,\\ G + G \to a&, \quad G + G \to a + j,
\end{aligned}
\end{equation}
where $V$ is a $U_{X}(1)$ mediator (dark photons, $B-L$, \dots), and $j$ is parton.

For the DIS production processes, we choose the invariant mass of the quark-antiquark pair as the scale of the process (\texttt{dynamical\_scale\_choice = 4}). Although \texttt{SensCalc} already includes the tabulated angle-energy distributions of the FIPs produced by DIS, it also includes the UFO files, allowing the user to re-generate these distributions under different assumptions if needed.

For the FIP decay processes, we extract the phase space of the metastable decay products for several FIP masses, select the sets of decay products that occur most frequently for the given decay, and export them in a format suitable for \texttt{Mathematica}. Interpolating the resulting phase space as a function of the FIP mass, we may then use it to compute the decay products acceptance, similarly to ordinary FIP decays for which it is possible to write analytical matrix elements.

\begin{table}[!h]
    \centering
    \begin{tabular}{|c|c|c|c|c|c|}
       \hline Particle & Fermilab (dump) & SPS & LHC & FCC-hh \\ \hline
       $\pi^{0}/\eta/\eta'/\rho^{0}/\omega/\gamma$  & \cite{Dobrich:2019dxc} & \cite{Dobrich:2019dxc} & \cite{Pierog:2013ria} & \cite{Pierog:2013ria} \\ \hline
        $B,D$  & \cite{Jerhot:2022chi} & \cite{CERN-SHiP-NOTE-2015-009} & \cite{Kling:2021fwx} & \cite{Kling:2021fwx}  \\ \hline
         $W,h,Z$   & -- & -- & \cite{Kling:2021fwx} & \cite{Kling:2021fwx} \\ \hline
    \end{tabular}
    \caption{List of the references used to generate, or directly take, the distributions of secondary particles that may produce FIPs.}
    \label{tab:secondary-particles-spectra}
\end{table}

The DIS production suffers from significant theoretical uncertainties. First, the choice of scale becomes important for light FIPs with masses $m_{\text{FIP}}\simeq 1-2\text{ GeV}$, where the uncertainties in the production cross-section may become $\mathcal{O}(1)$. Second, the minimal parton energy fraction required to produce a FIP is $x_{\text{min}} = m_{\text{FIP}}^{2}/s_{\text{pp}}$. For experiments like the LHC/FCC-hh and GeV-scale FIPs, $x_{\text{min}}$ can be as tiny as $10^{-8}$; this domain is only explored experimentally and is therefore subject to theoretical uncertainties (see Ref.~\cite{Berlin:2018jbm}). This becomes especially problematic in the case of the FCC-hh. Because of this, we do not consider the DIS production channel for the FCC-hh-based experiments.

\subsection{Production by secondary particles}
\label{app:secondary-particles-references}
Another important FIP production mechanism is through secondary particles --- either in their decays or scatterings. We handle this case by either generating the distributions of secondary particles using existing approaches from the literature, or directly using pre-calculated distributions. The list of references is provided in Table~\ref{tab:secondary-particles-spectra}.

Typically, the production probability of the FIP from a parent particle $X$ is the same as from the anti-particle $\bar{X}$. For example, the probability of producing an HNL in decays of $D_{s}$ meson is the same as in decays of $\bar{D}_{s}$. Therefore, the total flux of FIPs from $X,\bar{X}$ is proportional to the sum of the fluxes of these particles, $(N_{X}f_{X}+N_{\bar{X}}f_{\bar{X}})$, where $N_{X}$ is the total number of produced $X$, and $f_{X}$ is the normalized distribution. Instead of providing separate distributions $f_{X},f_{\bar{X}}$, we compute the weighted sum
\begin{equation}
    f_{X,\bar{X}} = \frac{N_{X}f_{X}+N_{\bar{X}}f_{\bar{X}}}{N_{X}+N_{\bar{X}}}
\end{equation}
For particles such as $B$ and $D$ mesons, $N_{X} = N_{\bar{X}}$ since the parent quarks $c,b$ are always produced together with their corresponding antiquarks. However, the shape of their distributions may be different. For particles such as $W$ bosons, not only the shape but also the numbers~$N$ are different, since the production processes of $W^\pm$ differ.

\section{\texttt{SensMC}: a simplified Monte-Carlo used for validation}
\label{app:toy-mc}
As an additional cross-check of \texttt{SensCalc}, we have implemented \texttt{SensMC}~\cite{SensMC-GitHub}, a small, customizable weight-based Monte-Carlo simulation, as an alternative way of numerically integrating Eq.~\eqref{eq:Nevents} for FIPs produced in meson decays. It makes extensive use of importance sampling in order to handle the (typically tiny) branching ratios of mesons to FIPs and the (possibly very displaced) decay vertex of the FIP. \texttt{SensMC} is written in the \texttt{Julia} programming language~\cite{julialang} in order to combine performance and readability, and it is released alongside \texttt{SensCalc} in the same repository~\cite{SensCalc-Zenodo}, as well as on GitHub.\footnote{The GitHub repository can be found at \url{https://github.com/JLTastet/SensMC}\,.}

\texttt{SensMC} numerically estimates Eq.~\eqref{eq:Nevents} using Monte-Carlo integration with importance sampling, by randomly generating a large number of weighted samples whose expectation values are $N_{\mathrm{ev}}$, and finally averaging them. The value of each random sample is computed as follows:
\begin{enumerate}
\item A meson species is randomly sampled based on the proportion of produced mesons of this species, with the event weight initially set to the total number of mesons produced across all species. The meson momentum is then randomly sampled from a precomputed spectrum (either a list for the spectrums from \texttt{FairShip}~\cite{SHiP:2018xqw} or a grid for those from \texttt{FORESEE}~\cite{Kling:2021fwx}). To account for potential variations in the atomic weight of the target, that would affect the overall normalization of the spectrums, the event is optionally reweighted using the formula $w_A = w_{\mathrm{Mo}} (A/96)^{0.29}$~\cite{Carvalho:2003pza}, with $A$ denoting the atomic weight of the target and assuming that the spectrums were initially computed for a molybdenum target (as is the case for the \texttt{FairShip} spectrums).

\item The FIP production channel is randomly selected from the decays of the parent meson, with a probability proportional to its branching ratio, and the event is reweighted by the total branching ratio to FIPs of the parent meson. Upon the meson decay, the momenta of its decay products, including the FIP, are uniformly sampled in phase space. The present simulation currently does not take into account the matrix elements because it cannot compute them all, however the logic needed to handle them is already present, allowing the user to implement their own matrix elements if needed.

\item The FIP's decay vertex is then selected randomly along its trajectory by either a)~sampling the proper lifetime from an exponential distribution and calculating the corresponding distance in the lab frame or b)~employing importance sampling, which restricts the position of the decay vertex to a shell covering the full decay volume, and then reweights the event by the ratio of the true decay distribution to the importance distribution. The FIP decay mode is selected similarly to its production mode, with a sampling probability proportional (and in most cases equal) to its branching ratio; and the event is reweighted by the total branching ratio of the implemented channels. The momenta of the FIP decay products are uniformly sampled in phase space in the current version (but matrix elements could in principle be taken into account, just like for the FIP production).

\item Following a similar procedure, any unstable Standard Model particles are recursively decayed until only metastable particles (that live long enough to be detected) remain, assuming the branching ratios listed in the \texttt{particletools} Python package. The acceptance condition is then evaluated on the set of final metastable particles produced in the FIP decay. The event weight is recorded, along with whether the event is accepted or not.
\end{enumerate}

Because each event is initially weighted by the total number of mesons, all event weights must finally be divided by the number of generated events. The sum of all weights then provides a numerical estimate of the total number of physical events (with the FIP decay vertex within the "shell" in case importance sampling is used), while the sum of event weights multiplied by their corresponding (binary) acceptances gives the total number of accepted events; the latter is independent of the specific importance distribution, as long as it fully covers the decay volume.

The sensitivity curve is computed iteratively, starting from a coarse grid in $(\log(m),\allowbreak \log(\theta))$ that covers the region where the experiment is susceptible to be sensitive. The expected number of accepted events is computed at each grid point. The multi-dimensional bisection method (MDBM)~\cite{MDBM} is then used to iteratively refine the grid in the vicinity of the iso-contour corresponding to (for example) 2.3 accepted events (for an exclusion sensitivity at the 90\% confidence level), effectively bisecting it without the need to evaluate a dense grid, which would be computationally costly. The final curve is then obtained from bilinear interpolation of the sparse grid values.

\newpage

\bibliographystyle{utphys}
\bibliography{bib.bib}

\providecommand{\href}[2]{#2}\begingroup\raggedright\begin{thebibliography}{10}

\bibitem{Beacham:2019nyx}
J.~Beacham {\em et~al.}, ``{Physics Beyond Colliders at CERN: Beyond the
  Standard Model Working Group Report},''
  \href{https://dx.doi.org/10.1088/1361-6471/ab4cd2}{{\em J. Phys. G}
  {\bfseries 47} no.~1, (2020) 010501},
  \href{https://arxiv.org/abs/1901.09966}{{\ttfamily arXiv:1901.09966
  [hep-ex]}}.

\bibitem{Alekhin:2015byh}
S.~Alekhin {\em et~al.}, ``{A facility to Search for Hidden Particles at the
  CERN SPS: the SHiP physics case},''
  \href{https://dx.doi.org/10.1088/0034-4885/79/12/124201}{{\em Rept. Prog.
  Phys.} {\bfseries 79} no.~12, (2016) 124201},
  \href{https://arxiv.org/abs/1504.04855}{{\ttfamily arXiv:1504.04855
  [hep-ph]}}.

\bibitem{Aberle:2839677}
{\bfseries SHiP} Collaboration, O.~Aberle {\em et~al.}, ``{BDF/SHiP at the ECN3
  high-intensity beam facility},'' tech. rep., CERN, Geneva, 2022.
\newblock \url{http://cds.cern.ch/record/2839677}.

\bibitem{Alviggi:2839484}
{\bfseries SHADOWS} Collaboration, M.~Alviggi {\em et~al.}, ``{SHADOWS Letter
  of Intent},'' tech. rep., CERN, Geneva, 2022.
\newblock \url{https://cds.cern.ch/record/2839484}.

\bibitem{CortinaGil:2839661}
{\bfseries HIKE} Collaboration, E.~Cortina~Gil {\em et~al.}, ``{HIKE, High
  Intensity Kaon Experiments at the CERN SPS: Letter of Intent},'' tech. rep.,
  CERN, Geneva, 2022.
\newblock \href{https://arxiv.org/abs/2211.16586}{{\ttfamily
  arXiv:2211.16586}}.
\newblock \url{https://cds.cern.ch/record/2839661}.
\newblock Letter of Intent submitted to CERN SPSC. Address all correspondence
  to hike-eb@cern.ch.

\bibitem{DUNE:2020ypp}
{\bfseries DUNE} Collaboration, B.~Abi {\em et~al.}, ``{Deep Underground
  Neutrino Experiment (DUNE), Far Detector Technical Design Report, Volume II:
  DUNE Physics},'' \href{https://arxiv.org/abs/2002.03005}{{\ttfamily
  arXiv:2002.03005 [hep-ex]}}.

\bibitem{DUNE:2020fgq}
{\bfseries DUNE} Collaboration, B.~Abi {\em et~al.}, ``{Prospects for beyond
  the Standard Model physics searches at the Deep Underground Neutrino
  Experiment},'' \href{https://dx.doi.org/10.1140/epjc/s10052-021-09007-w}{{\em
  Eur. Phys. J. C} {\bfseries 81} no.~4, (2021) 322},
  \href{https://arxiv.org/abs/2008.12769}{{\ttfamily arXiv:2008.12769
  [hep-ex]}}.

\bibitem{Batell:2020vqn}
B.~Batell, J.~A. Evans, S.~Gori, and M.~Rai, ``{Dark Scalars and Heavy Neutral
  Leptons at DarkQuest},''
  \href{https://dx.doi.org/10.1007/JHEP05(2021)049}{{\em JHEP} {\bfseries 05}
  (2021) 049}, \href{https://arxiv.org/abs/2008.08108}{{\ttfamily
  arXiv:2008.08108 [hep-ph]}}.

\bibitem{MATHUSLA:2019qpy}
{\bfseries MATHUSLA} Collaboration, H.~Lubatti {\em et~al.}, ``{Explore the
  lifetime frontier with MATHUSLA},''
  \href{https://dx.doi.org/10.1088/1748-0221/15/06/C06026}{{\em JINST}
  {\bfseries 15} no.~06, (2020) C06026},
  \href{https://arxiv.org/abs/1901.04040}{{\ttfamily arXiv:1901.04040
  [hep-ex]}}.

\bibitem{Cerci:2021nlb}
S.~Cerci {\em et~al.}, ``{FACET: A new long-lived particle detector in the very
  forward region of the CMS experiment},''
  \href{https://dx.doi.org/10.1007/JHEP06(2022)110}{{\em JHEP} {\bfseries 2022}
  no.~06, (2022) 110}, \href{https://arxiv.org/abs/2201.00019}{{\ttfamily
  arXiv:2201.00019 [hep-ex]}}.

\bibitem{FASER:2018bac}
{\bfseries FASER} Collaboration, A.~Ariga {\em et~al.}, ``{Technical Proposal
  for FASER: ForwArd Search ExpeRiment at the LHC},''
  \href{https://arxiv.org/abs/1812.09139}{{\ttfamily arXiv:1812.09139
  [physics.ins-det]}}.

\bibitem{SHiP:2020sos}
{\bfseries SHiP} Collaboration, C.~Ahdida {\em et~al.}, ``{SND@LHC},''
  \href{https://arxiv.org/abs/2002.08722}{{\ttfamily arXiv:2002.08722
  [physics.ins-det]}}.

\bibitem{Bauer:2019vqk}
M.~Bauer, O.~Brandt, L.~Lee, and C.~Ohm, ``{ANUBIS: Proposal to search for
  long-lived neutral particles in CERN service shafts},''
  \href{https://arxiv.org/abs/1909.13022}{{\ttfamily arXiv:1909.13022
  [physics.ins-det]}}.

\bibitem{Aielli:2019ivi}
G.~Aielli {\em et~al.}, ``{Expression of interest for the CODEX-b detector},''
  \href{https://dx.doi.org/10.1140/epjc/s10052-020-08711-3}{{\em Eur. Phys. J.
  C} {\bfseries 80} no.~12, (2020) 1177},
  \href{https://arxiv.org/abs/1911.00481}{{\ttfamily arXiv:1911.00481
  [hep-ex]}}.

\bibitem{Dercks:2018wum}
D.~Dercks, H.~K. Dreiner, M.~Hirsch, and Z.~S. Wang, ``{Long-Lived Fermions at
  AL3X},'' \href{https://dx.doi.org/10.1103/PhysRevD.99.055020}{{\em Phys. Rev.
  D} {\bfseries 99} no.~5, (2019) 055020},
  \href{https://arxiv.org/abs/1811.01995}{{\ttfamily arXiv:1811.01995
  [hep-ph]}}.

\bibitem{Boyarsky:2022epg}
A.~Boyarsky, O.~Mikulenko, M.~Ovchynnikov, and L.~Shchutska, ``{Exploring the
  potential of FCC-hh to search for particles from B mesons},''
  \href{https://dx.doi.org/10.1007/JHEP01(2023)042}{{\em JHEP} {\bfseries 01}
  (2023) 042}, \href{https://arxiv.org/abs/2204.01622}{{\ttfamily
  arXiv:2204.01622 [hep-ph]}}.

\bibitem{Kling:2021fwx}
F.~Kling and S.~Trojanowski, ``{Forward experiment sensitivity estimator for
  the LHC and future hadron colliders},''
  \href{https://dx.doi.org/10.1103/PhysRevD.104.035012}{{\em Phys. Rev. D}
  {\bfseries 104} no.~3, (2021) 035012},
  \href{https://arxiv.org/abs/2105.07077}{{\ttfamily arXiv:2105.07077
  [hep-ph]}}.

\bibitem{Jerhot:2022chi}
J.~Jerhot, B.~D\"obrich, F.~Ertas, F.~Kahlhoefer, and T.~Spadaro, ``{ALPINIST:
  Axion-Like Particles In Numerous Interactions Simulated and Tabulated},''
  \href{https://dx.doi.org/10.1007/JHEP07(2022)094}{{\em JHEP} {\bfseries 07}
  (2022) 094}, \href{https://arxiv.org/abs/2201.05170}{{\ttfamily
  arXiv:2201.05170 [hep-ph]}}.

\bibitem{Mathematica}
{Wolfram Research, Inc.}, ``Mathematica, {V}ersion 13.2,''
\newblock \url{https://www.wolfram.com/mathematica}. Champaign, IL, 2022.

\bibitem{SensCalc-Zenodo}
M.~Ovchynnikov, ``{SensCalc}.'' 2023.
\newblock \url{https://doi.org/10.5281/zenodo.7957784}.

\bibitem{Bondarenko:2019yob}
K.~Bondarenko, A.~Boyarsky, M.~Ovchynnikov, and O.~Ruchayskiy, ``{Sensitivity
  of the intensity frontier experiments for neutrino and scalar portals:
  analytic estimates},'' \href{https://dx.doi.org/10.1007/JHEP08(2019)061}{{\em
  JHEP} {\bfseries 08} (2019) 061},
  \href{https://arxiv.org/abs/1902.06240}{{\ttfamily arXiv:1902.06240
  [hep-ph]}}.

\bibitem{Boiarska:2021yho}
I.~Boiarska, A.~Boyarsky, O.~Mikulenko, and M.~Ovchynnikov, ``{Constraints from
  the CHARM experiment on heavy neutral leptons with tau mixing},''
  \href{https://dx.doi.org/10.1103/PhysRevD.104.095019}{{\em Phys. Rev. D}
  {\bfseries 104} no.~9, (2021) 095019},
  \href{https://arxiv.org/abs/2107.14685}{{\ttfamily arXiv:2107.14685
  [hep-ph]}}.

\bibitem{Ovchynnikov:2022its}
M.~Ovchynnikov, V.~Kryshtal, and K.~Bondarenko, ``{Sensitivity of the FACET
  experiment to Heavy Neutral Leptons and Dark Scalars},''
  \href{https://dx.doi.org/10.1007/JHEP02(2023)056}{{\em JHEP} {\bfseries 02}
  (2023) 056}, \href{https://arxiv.org/abs/2209.14870}{{\ttfamily
  arXiv:2209.14870 [hep-ph]}}.

\bibitem{Coloma:2023adi}
P.~Coloma, J.~L\'opez-Pav\'on, L.~Molina-Bueno, and S.~Urrea, ``{New Physics
  searches using ProtoDUNE and the CERN SPS accelerator},''
  \href{https://arxiv.org/abs/2304.06765}{{\ttfamily arXiv:2304.06765
  [hep-ph]}}.

\bibitem{Batell:2023mdn}
B.~Batell, W.~Huang, and K.~J. Kelly, ``{Keeping it simple: simplified
  frameworks for long-lived particles at neutrino facilities},''
  \href{https://dx.doi.org/10.1007/JHEP08(2023)092}{{\em JHEP} {\bfseries 08}
  (2023) 092}, \href{https://arxiv.org/abs/2304.11189}{{\ttfamily
  arXiv:2304.11189 [hep-ph]}}.

\bibitem{SensMC-GitHub}
J.-L. Tastet, ``{SensMC}.'' 2023.
\newblock \url{https://github.com/JLTastet/SensMC}.

\bibitem{Bondarenko:2023fex}
K.~Bondarenko, A.~Boyarsky, O.~Mikulenko, R.~Jacobsson, and M.~Ovchynnikov,
  ``{Towards the optimal beam dump experiment to search for feebly interacting
  particles},'' \href{https://arxiv.org/abs/2304.02511}{{\ttfamily
  arXiv:2304.02511 [hep-ph]}}.

\bibitem{SHiP:2018xqw}
{\bfseries SHiP} Collaboration, C.~Ahdida {\em et~al.}, ``{Sensitivity of the
  SHiP experiment to Heavy Neutral Leptons},''
  \href{https://dx.doi.org/10.1007/JHEP04(2019)077}{{\em JHEP} {\bfseries 04}
  (2019) 077}, \href{https://arxiv.org/abs/1811.00930}{{\ttfamily
  arXiv:1811.00930 [hep-ph]}}.

\bibitem{SHiP:2020vbd}
{\bfseries SHiP} Collaboration, C.~Ahdida {\em et~al.}, ``{Sensitivity of the
  SHiP experiment to dark photons decaying to a pair of charged particles},''
  \href{https://dx.doi.org/10.1140/epjc/s10052-021-09224-3}{{\em Eur. Phys. J.
  C} {\bfseries 81} no.~5, (2021) 451},
  \href{https://arxiv.org/abs/2011.05115}{{\ttfamily arXiv:2011.05115
  [hep-ex]}}.

\bibitem{CERN-SHiP-NOTE-2015-009}
{\bfseries SHiP} Collaboration, H.~Dijkstra and T.~Ruf, ``{Heavy Flavour
  Cascade Production in a Beam Dump}.'' Dec, 2015.
\newblock \url{https://cds.cern.ch/record/2115534}. CERN-SHiP-NOTE-2015-009.

\bibitem{Boiarska:2019jym}
I.~Boiarska, K.~Bondarenko, A.~Boyarsky, V.~Gorkavenko, M.~Ovchynnikov, and
  A.~Sokolenko, ``{Phenomenology of GeV-scale scalar portal},''
  \href{https://dx.doi.org/10.1007/JHEP11(2019)162}{{\em JHEP} {\bfseries 11}
  (2019) 162}, \href{https://arxiv.org/abs/1904.10447}{{\ttfamily
  arXiv:1904.10447 [hep-ph]}}.

\bibitem{MATHUSLA:2022sze}
{\bfseries MATHUSLA} Collaboration, C.~Alpigiani {\em et~al.}, ``{Recent
  Progress and Next Steps for the MATHUSLA LLP Detector},'' in {\em {2022
  Snowmass Summer Study}}.
\newblock 3, 2022.
\newblock \href{https://arxiv.org/abs/2203.08126}{{\ttfamily arXiv:2203.08126
  [hep-ex]}}.

\bibitem{Harland-Lang:2019zur}
L.~Harland-Lang, J.~Jaeckel, and M.~Spannowsky, ``{A fresh look at ALP searches
  in fixed target experiments},''
  \href{https://dx.doi.org/10.1016/j.physletb.2019.04.045}{{\em Phys. Lett. B}
  {\bfseries 793} (2019) 281--289},
  \href{https://arxiv.org/abs/1902.04878}{{\ttfamily arXiv:1902.04878
  [hep-ph]}}.

\bibitem{Buonocore:2018xjk}
L.~Buonocore, C.~Frugiuele, F.~Maltoni, O.~Mattelaer, and F.~Tramontano,
  ``{Event generation for beam dump experiments},''
  \href{https://dx.doi.org/10.1007/JHEP05(2019)028}{{\em JHEP} {\bfseries 05}
  (2019) 028}, \href{https://arxiv.org/abs/1812.06771}{{\ttfamily
  arXiv:1812.06771 [hep-ph]}}.

\bibitem{deNiverville:2016rqh}
P.~deNiverville, C.-Y. Chen, M.~Pospelov, and A.~Ritz, ``{Light dark matter in
  neutrino beams: production modelling and scattering signatures at MiniBooNE,
  T2K and SHiP},'' \href{https://dx.doi.org/10.1103/PhysRevD.95.035006}{{\em
  Phys. Rev. D} {\bfseries 95} no.~3, (2017) 035006},
  \href{https://arxiv.org/abs/1609.01770}{{\ttfamily arXiv:1609.01770
  [hep-ph]}}.

\bibitem{Ovchynnikov:2022rqj}
M.~Ovchynnikov, T.~Schwetz, and J.-Y. Zhu, ``{Dipole portal and neutrinophilic
  scalars at DUNE revisited: The importance of the high-energy neutrino
  tail},'' \href{https://dx.doi.org/10.1103/PhysRevD.107.055029}{{\em Phys.
  Rev. D} {\bfseries 107} no.~5, (2023) 055029},
  \href{https://arxiv.org/abs/2210.13141}{{\ttfamily arXiv:2210.13141
  [hep-ph]}}.

\bibitem{Ballett:2019bgd}
P.~Ballett, T.~Boschi, and S.~Pascoli, ``{Heavy Neutral Leptons from low-scale
  seesaws at the DUNE Near Detector},''
  \href{https://dx.doi.org/10.1007/JHEP03(2020)111}{{\em JHEP} {\bfseries 03}
  (2020) 111}, \href{https://arxiv.org/abs/1905.00284}{{\ttfamily
  arXiv:1905.00284 [hep-ph]}}.

\bibitem{Kling:2022ehv}
F.~Kling and P.~Qu\'\i{}lez, ``{ALP searches at the LHC: FASER as a
  light-shining-through-walls experiment},''
  \href{https://dx.doi.org/10.1103/PhysRevD.106.055036}{{\em Phys. Rev. D}
  {\bfseries 106} no.~5, (2022) 055036},
  \href{https://arxiv.org/abs/2204.03599}{{\ttfamily arXiv:2204.03599
  [hep-ph]}}.

\bibitem{Ahdida:2867743}
C.~Ahdida {\em et~al.}, ``{Post-LS3 Experimental Options in ECN3},'' tech.
  rep., CERN, Geneva, 2023.
\newblock \url{https://cds.cern.ch/record/2867743}.

\bibitem{DUNE:2021tad}
{\bfseries DUNE} Collaboration, V.~Hewes {\em et~al.}, ``{Deep Underground
  Neutrino Experiment (DUNE) Near Detector Conceptual Design Report},''
  \href{https://dx.doi.org/10.3390/instruments5040031}{{\em Instruments}
  {\bfseries 5} no.~4, (2021) 31},
  \href{https://arxiv.org/abs/2103.13910}{{\ttfamily arXiv:2103.13910
  [physics.ins-det]}}.

\bibitem{FASER:2019aik}
{\bfseries FASER} Collaboration, A.~Ariga {\em et~al.}, ``{FASER: ForwArd
  Search ExpeRiment at the LHC},''
  \href{https://arxiv.org/abs/1901.04468}{{\ttfamily arXiv:1901.04468
  [hep-ex]}}.

\bibitem{FASER:2020gpr}
{\bfseries FASER} Collaboration, H.~Abreu {\em et~al.}, ``{Technical Proposal:
  FASERnu},'' \href{https://arxiv.org/abs/2001.03073}{{\ttfamily
  arXiv:2001.03073 [physics.ins-det]}}.

\bibitem{Feng:2022inv}
J.~L. Feng {\em et~al.}, ``{The Forward Physics Facility at the High-Luminosity
  LHC},'' \href{https://dx.doi.org/10.1088/1361-6471/ac865e}{{\em J. Phys. G}
  {\bfseries 50} no.~3, (2023) 030501},
  \href{https://arxiv.org/abs/2203.05090}{{\ttfamily arXiv:2203.05090
  [hep-ex]}}.

\bibitem{CHARM:1983ayi}
{\bfseries CHARM} Collaboration, F.~Bergsma {\em et~al.}, ``{A Search for
  Decays of Heavy Neutrinos},''
  \href{https://dx.doi.org/10.1016/0370-2693(83)90275-7}{{\em Phys. Lett. B}
  {\bfseries 128} (1983) 361}.

\bibitem{BEBCWA66:1986err}
{\bfseries BEBC WA66} Collaboration, H.~Grassler {\em et~al.}, ``{Prompt
  Neutrino Production in 400-{GeV} Proton Copper Interactions},''
  \href{https://dx.doi.org/10.1016/0550-3213(86)90246-4}{{\em Nucl. Phys. B}
  {\bfseries 273} (1986) 253--274}.

\bibitem{Ilten:2018crw}
P.~Ilten, Y.~Soreq, M.~Williams, and W.~Xue, ``{Serendipity in dark photon
  searches},'' \href{https://dx.doi.org/10.1007/JHEP06(2018)004}{{\em JHEP}
  {\bfseries 06} (2018) 004},
  \href{https://arxiv.org/abs/1801.04847}{{\ttfamily arXiv:1801.04847
  [hep-ph]}}.

\bibitem{Bondarenko:2018ptm}
K.~Bondarenko, A.~Boyarsky, D.~Gorbunov, and O.~Ruchayskiy, ``{Phenomenology of
  GeV-scale Heavy Neutral Leptons},''
  \href{https://dx.doi.org/10.1007/JHEP11(2018)032}{{\em JHEP} {\bfseries 11}
  (2018) 032}, \href{https://arxiv.org/abs/1805.08567}{{\ttfamily
  arXiv:1805.08567 [hep-ph]}}.

\bibitem{Dobrich:2019dxc}
B.~D\"obrich, J.~Jaeckel, and T.~Spadaro, ``{Light in the beam dump - ALP
  production from decay photons in proton beam-dumps},''
  \href{https://dx.doi.org/10.1007/JHEP05(2019)213}{{\em JHEP} {\bfseries 05}
  (2019) 213}, \href{https://arxiv.org/abs/1904.02091}{{\ttfamily
  arXiv:1904.02091 [hep-ph]}}. [Erratum: JHEP 10, 046 (2020)].

\bibitem{DallaValleGarcia:2023xhh}
G.~Dalla Valle~Garcia, F.~Kahlhoefer, M.~Ovchynnikov, and A.~Zaporozhchenko,
  ``{Phenomenology of axion-like particles with universal fermion couplings --
  revisited},'' \href{https://arxiv.org/abs/2310.03524}{{\ttfamily
  arXiv:2310.03524 [hep-ph]}}.

\bibitem{Aloni:2018vki}
D.~Aloni, Y.~Soreq, and M.~Williams, ``{Coupling QCD-Scale Axionlike Particles
  to Gluons},'' \href{https://dx.doi.org/10.1103/PhysRevLett.123.031803}{{\em
  Phys. Rev. Lett.} {\bfseries 123} no.~3, (2019) 031803},
  \href{https://arxiv.org/abs/1811.03474}{{\ttfamily arXiv:1811.03474
  [hep-ph]}}.

\bibitem{Chakraborty:2021wda}
S.~Chakraborty, M.~Kraus, V.~Loladze, T.~Okui, and K.~Tobioka, ``{Heavy QCD
  axion in b\textrightarrow{}s transition: Enhanced limits and projections},''
  \href{https://dx.doi.org/10.1103/PhysRevD.104.055036}{{\em Phys. Rev. D}
  {\bfseries 104} no.~5, (2021) 055036},
  \href{https://arxiv.org/abs/2102.04474}{{\ttfamily arXiv:2102.04474
  [hep-ph]}}.

\bibitem{Tulin:2014tya}
S.~Tulin, ``{New weakly-coupled forces hidden in low-energy QCD},''
  \href{https://dx.doi.org/10.1103/PhysRevD.89.114008}{{\em Phys. Rev. D}
  {\bfseries 89} no.~11, (2014) 114008},
  \href{https://arxiv.org/abs/1404.4370}{{\ttfamily arXiv:1404.4370 [hep-ph]}}.

\bibitem{Alwall:2014hca}
J.~Alwall, R.~Frederix, S.~Frixione, V.~Hirschi, F.~Maltoni, O.~Mattelaer,
  H.~S. Shao, T.~Stelzer, P.~Torrielli, and M.~Zaro, ``{The automated
  computation of tree-level and next-to-leading order differential cross
  sections, and their matching to parton shower simulations},''
  \href{https://dx.doi.org/10.1007/JHEP07(2014)079}{{\em JHEP} {\bfseries 07}
  (2014) 079}, \href{https://arxiv.org/abs/1405.0301}{{\ttfamily
  arXiv:1405.0301 [hep-ph]}}.

\bibitem{Alloul:2013bka}
A.~Alloul, N.~D. Christensen, C.~Degrande, C.~Duhr, and B.~Fuks, ``{FeynRules
  2.0 - A complete toolbox for tree-level phenomenology},''
  \href{https://dx.doi.org/10.1016/j.cpc.2014.04.012}{{\em Comput. Phys.
  Commun.} {\bfseries 185} (2014) 2250--2300},
  \href{https://arxiv.org/abs/1310.1921}{{\ttfamily arXiv:1310.1921 [hep-ph]}}.

\bibitem{Degrande:2011ua}
C.~Degrande, C.~Duhr, B.~Fuks, D.~Grellscheid, O.~Mattelaer, and T.~Reiter,
  ``{UFO - The Universal FeynRules Output},''
  \href{https://dx.doi.org/10.1016/j.cpc.2012.01.022}{{\em Comput. Phys.
  Commun.} {\bfseries 183} (2012) 1201--1214},
  \href{https://arxiv.org/abs/1108.2040}{{\ttfamily arXiv:1108.2040 [hep-ph]}}.

\bibitem{Sjostrand:2014zea}
T.~Sj\"ostrand, S.~Ask, J.~R. Christiansen, R.~Corke, N.~Desai, P.~Ilten,
  S.~Mrenna, S.~Prestel, C.~O. Rasmussen, and P.~Z. Skands, ``{An introduction
  to PYTHIA 8.2}'' \href{https://dx.doi.org/10.1016/j.cpc.2015.01.024}{{\em
  Comput. Phys. Commun.} {\bfseries 191} (2015) 159--177},
  \href{https://arxiv.org/abs/1410.3012}{{\ttfamily arXiv:1410.3012 [hep-ph]}}.

\bibitem{Winkler:2018qyg}
M.~W. Winkler, ``{Decay and detection of a light scalar boson mixing with the
  Higgs boson},'' \href{https://dx.doi.org/10.1103/PhysRevD.99.015018}{{\em
  Phys. Rev. D} {\bfseries 99} no.~1, (2019) 015018},
  \href{https://arxiv.org/abs/1809.01876}{{\ttfamily arXiv:1809.01876
  [hep-ph]}}.

\bibitem{Spira:1995rr}
M.~Spira, A.~Djouadi, D.~Graudenz, and P.~M. Zerwas, ``{Higgs boson production
  at the LHC},'' \href{https://dx.doi.org/10.1016/0550-3213(95)00379-7}{{\em
  Nucl. Phys. B} {\bfseries 453} (1995) 17--82},
  \href{https://arxiv.org/abs/hep-ph/9504378}{{\ttfamily
  arXiv:hep-ph/9504378}}.

\bibitem{Carvalho:2003pza}
J.~Carvalho, ``{Compilation of cross sections for proton nucleus interactions
  at the HERA energy},''
  \href{https://dx.doi.org/10.1016/S0375-9474(03)01597-5}{{\em Nucl. Phys. A}
  {\bfseries 725} (2003) 269--275}.

\bibitem{Shtabovenko:2020gxv}
V.~Shtabovenko, R.~Mertig, and F.~Orellana, ``{FeynCalc 9.3: New features and
  improvements},'' \href{https://dx.doi.org/10.1016/j.cpc.2020.107478}{{\em
  Comput. Phys. Commun.} {\bfseries 256} (2020) 107478},
  \href{https://arxiv.org/abs/2001.04407}{{\ttfamily arXiv:2001.04407
  [hep-ph]}}.

\bibitem{Tastet:2021vwp}
J.-L. Tastet, O.~Ruchayskiy, and I.~Timiryasov, ``{Reinterpreting the ATLAS
  bounds on heavy neutral leptons in a realistic neutrino oscillation model},''
  \href{https://dx.doi.org/10.1007/JHEP12(2021)182}{{\em JHEP} {\bfseries 12}
  (2021) 182}, \href{https://arxiv.org/abs/2107.12980}{{\ttfamily
  arXiv:2107.12980 [hep-ph]}}.

\bibitem{Abada:2022wvh}
A.~Abada, P.~Escribano, X.~Marcano, and G.~Piazza, ``{Collider searches for
  heavy neutral leptons: beyond simplified scenarios},''
  \href{https://dx.doi.org/10.1140/epjc/s10052-022-11011-7}{{\em Eur. Phys. J.
  C} {\bfseries 82} no.~11, (2022) 1030},
  \href{https://arxiv.org/abs/2208.13882}{{\ttfamily arXiv:2208.13882
  [hep-ph]}}.

\bibitem{Beltran:2023nli}
R.~Beltr\'an, G.~Cottin, M.~Hirsch, A.~Titov, and Z.~S. Wang,
  ``{Reinterpretation of searches for long-lived particles from meson
  decays},'' \href{https://dx.doi.org/10.1007/JHEP05(2023)031}{{\em JHEP}
  {\bfseries 05} (2023) 031},
  \href{https://arxiv.org/abs/2302.03216}{{\ttfamily arXiv:2302.03216
  [hep-ph]}}.

\bibitem{Antel:2023hkf}
C.~Antel {\em et~al.}, ``{Feebly Interacting Particles: FIPs 2022 workshop
  report},'' \href{https://arxiv.org/abs/2305.01715}{{\ttfamily
  arXiv:2305.01715 [hep-ph]}}.

\bibitem{Boiarska:2019jcw}
I.~Boiarska, K.~Bondarenko, A.~Boyarsky, S.~Eijima, M.~Ovchynnikov,
  O.~Ruchayskiy, and I.~Timiryasov, ``{Probing baryon asymmetry of the Universe
  at LHC and SHiP},'' \href{https://arxiv.org/abs/1902.04535}{{\ttfamily
  arXiv:1902.04535 [hep-ph]}}.

\bibitem{Monin:2018lee}
A.~Monin, A.~Boyarsky, and O.~Ruchayskiy, ``{Hadronic decays of a light
  Higgs-like scalar},''
  \href{https://dx.doi.org/10.1103/PhysRevD.99.015019}{{\em Phys. Rev. D}
  {\bfseries 99} no.~1, (2019) 015019},
  \href{https://arxiv.org/abs/1806.07759}{{\ttfamily arXiv:1806.07759
  [hep-ph]}}.

\bibitem{Gorbunov:2023lga}
D.~Gorbunov, E.~Kriukova, and O.~Teryaev, ``{Scalar decay into pions via Higgs
  portal},'' \href{https://arxiv.org/abs/2303.12847}{{\ttfamily
  arXiv:2303.12847 [hep-ph]}}.

\bibitem{Feng:2017vli}
J.~L. Feng, I.~Galon, F.~Kling, and S.~Trojanowski, ``{Dark Higgs bosons at the
  ForwArd Search ExpeRiment},''
  \href{https://dx.doi.org/10.1103/PhysRevD.97.055034}{{\em Phys. Rev. D}
  {\bfseries 97} no.~5, (2018) 055034},
  \href{https://arxiv.org/abs/1710.09387}{{\ttfamily arXiv:1710.09387
  [hep-ph]}}.

\bibitem{Bezrukov:2009yw}
F.~Bezrukov and D.~Gorbunov, ``{Light inflaton Hunter's Guide},''
  \href{https://dx.doi.org/10.1007/JHEP05(2010)010}{{\em JHEP} {\bfseries 05}
  (2010) 010}, \href{https://arxiv.org/abs/0912.0390}{{\ttfamily
  arXiv:0912.0390 [hep-ph]}}.

\bibitem{Pierog:2013ria}
T.~Pierog, I.~Karpenko, J.~M. Katzy, E.~Yatsenko, and K.~Werner, ``{EPOS LHC:
  Test of collective hadronization with data measured at the CERN Large Hadron
  Collider},'' \href{https://dx.doi.org/10.1103/PhysRevC.92.034906}{{\em Phys.
  Rev. C} {\bfseries 92} no.~3, (2015) 034906},
  \href{https://arxiv.org/abs/1306.0121}{{\ttfamily arXiv:1306.0121 [hep-ph]}}.

\bibitem{Berlin:2018jbm}
A.~Berlin and F.~Kling, ``{Inelastic Dark Matter at the LHC Lifetime Frontier:
  ATLAS, CMS, LHCb, CODEX-b, FASER, and MATHUSLA},''
  \href{https://dx.doi.org/10.1103/PhysRevD.99.015021}{{\em Phys. Rev. D}
  {\bfseries 99} no.~1, (2019) 015021},
  \href{https://arxiv.org/abs/1810.01879}{{\ttfamily arXiv:1810.01879
  [hep-ph]}}.

\bibitem{julialang}
J.~Bezanson, A.~Edelman, S.~Karpinski, and V.~B. Shah, ``Julia: A fresh
  approach to numerical computing,''
  \href{https://dx.doi.org/10.1137/141000671}{{\em SIAM Review} {\bfseries 59}
  no.~1, (2017) 65--98}. \url{https://doi.org/10.1137/141000671}.

\bibitem{MDBM}
D.~Bachrathy and G.~Stépán, ``Bisection method in higher dimensions and the
  efficiency number,'' \href{https://dx.doi.org/10.3311/pp.me.2012-2.01}{{\em
  Periodica Polytechnica Mechanical Engineering} {\bfseries 56} no.~2, (2012)
  81–86}. \url{https://pp.bme.hu/me/article/view/1236}.

\end{thebibliography}\endgroup

\end{document}